%% file: S_Channel_Simplified_Models.tex
\documentclass[pdftex,twocolumn,epjc3_preprint,runningheads]{svjour3}

\usepackage[colorlinks,citecolor=blue,urlcolor=blue,linkcolor=blue,breaklinks=true]{hyperref}
\usepackage[dvipsnames]{xcolor}

\input{R3_template}

\newcommand{\subparagraph}{} 
\journalname{Eur.\ Phys.\ J.\ C}
\smartqed

\usepackage[title]{appendix}

\usepackage{amssymb}
\usepackage{pifont}
\newcommand{\La}{{\rm \Lambda}}

\usepackage{natbib}
\usepackage{graphicx}
\usepackage{float}
\usepackage{scalerel}

\usepackage{multirow}
\usepackage{booktabs}
\usepackage{wrapfig}

\newcommand{\madanalysis}{\textsf{MadAnalysis}\xspace}

\newcommand{\MasterCode}{\textsf{MasterCode}\xspace}

\begin{document}

\title{Global fits of simplified models for dark matter with GAMBIT}
\subtitle{I. Scalar and fermionic models with $s$-channel vector mediators}

\author{Christopher Chang\thanksref{uq,a} \and
 Pat Scott\thanksref{qb} \and
 Tom\'as~E.~Gonzalo\thanksref{aachen,kitTTP} \and
 Felix Kahlhoefer\thanksref{aachen,kitTTP} \and
 Anders Kvellestad\thanksref{oslo} \and
 Martin White\thanksref{adelaide}
}

\institute{
  \gi{uq}
  \gi{qb}
  \gi{aachen}
  \gi{kitTTP}
  \gi{adelaide}
  \last{oslo}
}

\thankstext{a}{christopher.chang@uq.net.au}

\titlerunning{Global fits of simplified DM models I. Scalar and fermion DM with $s$-channel vector mediators}
\authorrunning{Chang et al.}

\date{Received: date / Accepted: date}

\preprintnumber{gambit-physics-2022, TTP22-060, P3H-22-098, ADP-22-29/T1200}

\maketitle

\begin{abstract}

Simplified models provide a useful way to study the impacts of a small number of new particles on experimental observables and the interplay of those observables, without the need to construct an underlying theory. In this study, we perform global fits of simplified dark matter models with \GB using an up-to-date set of likelihoods for indirect detection, direct detection and collider searches. We investigate models in which a scalar or fermionic dark matter candidate couples to quarks via an $s$-channel vector mediator. Large parts of parameter space survive for each model. In the case of Dirac or Majorana fermion dark matter, excesses in LHC monojet searches and relic density limits tend to prefer the resonance region, where the dark matter has approximately half the mass of the mediator. A combination of vector and axial-vector couplings to the Dirac candidate also leads to competing constraints from direct detection and unitarity violation.

\end{abstract}

\tableofcontents


\section{Introduction}

The Standard Model (SM) remains enormously successful as a theory of particle physics, but is widely thought to be incomplete and expected to be superseded by a more complete theory. One of the many motivations for searching for beyond-Standard Model (BSM) physics is to explain the dark matter (DM) evident in a number of astrophysical and cosmological observations \cite{Zwicky33, bullet, wmap3year}. The Weakly-Interacting Massive Particle (WIMP) hypothesis, in which DM is assumed to consist of a new species that interacts with a strength at least as weak as the weak nuclear force, is amongst the leading DM explanations due to its ability to explain the observed cosmological relic abundance of DM \cite{Lee:1977ua} at the same time as potentially being very tightly constrained in the near future by current experimental technologies \cite{Arcadi:2017kky}.

Whilst there are plenty of UV-complete theories that include viable WIMP candidates, it is advantageous to take a model-independent approach and construct a low-energy effective theory that includes the relevant phenomenology for our current experimental probes, whilst remaining agnostic about the high energy physics that we cannot currently observe. The simplest way to construct such a theory is to write down an \emph{effective field theory} (EFT), in which the SM Lagrangian density is extended with a number of effective operators that encode possible DM-SM interactions. An EFT is valid up to some scale $\La$, at which point one would start to resolve the physics that generates the operators in the low energy description. This typically makes EFTs useful for studying the impact of low energy experimental probes, such as indirect \cite{Goodman:2010qn, Beltran:2008xg, Cheung:2011nt, Harnik:2008uu, DeSimone:2013gj, Karwin:2016tsw} and direct \cite{Fan:2010gt,Agrawal:2010fh,Crivellin:2014gpa,Hoferichter:2016nvd,Kahlhoefer:2016eds} DM detection experiments, but difficult for higher-energy probes such as the ATLAS and CMS experiments at the Large Hadron Collider (LHC) \cite{Buchmueller:2013dya, Abdallah:2015ter, Abdallah:2014hon, Malik:2014ggr, Busoni:2013lha, Busoni:2014sya, Busoni:2014haa}. A detailed study of EFTs for Dirac DM has recently been performed by the \GB collaboration \cite{GAMBIT:2021rlp}.

An alternative, and only slightly more complicated, approach is to explicitly add a particle that mediates interactions between the DM candidate and SM species, giving what is popularly known as a \emph{simplified model}.  Most such models are themselves also EFTs. Previous reviews of such models can be found in Refs.~\cite{Abdallah:2015ter,Arina:2018zcq,DeSimone:2016fbz,Albert:2017onk,Boveia:2016mrp,Kahlhoefer:2017dnp,Arcadi:2017kky, Morgante:2018tiq}. In the limit of large mediator masses, the EFT approach may be recovered, but simplified models remain the preferred way to investigate the simultaneous impact of high energy collider and low energy DM probes \cite{DEramo:2016gos, Carpenter:2016thc, Abercrombie:2015wmb} on the landscape of possible DM-SM interactions.

There are a plethora of studies of simplified DM models in the literature, including most notably a global scan of an $s$-channel vector-mediated Dirac fermion DM model performed by the \MasterCode Collaboration \cite{Bagnaschi:2019djj}. In this paper, we perform global fits of $s$-channel vector-mediated scalar, Dirac and Majorana fermion DM models, using \GB \textsf{v2.3} \cite{gambit}. Setting up global fits of multiple simplified models in a single study has recently become more accessible with the tool for automatically generating \GB code, \gum \cite{GUM}. We complement the work of Ref.~\cite{Bagnaschi:2019djj} by also considering scalar and Majorana fermion DM, and Dirac fermion DM with both vector and axial vector DM couplings. The latter requires thorough treatment of direct detection signatures to allow concurrent contribution from spin-dependent and spin-indepent interactions.

Performing global fits of these models is of interest, due to the high degree of interplay between different experiments. In the case of Dirac DM, it has been shown in the EFT limit that requiring the relic abundance to be saturated can be incompatible with constraints from laboratory experiments for low DM masses \cite{GAMBIT:2021rlp}. We find that this incompatibility remains when the EFT is promoted to a simplified model. This lower bound on the DM mass is also present in the complex scalar DM model, where collider searches have very little influence on this bound. Suppressed direct detection signals in the Majorana DM model give no such lower bound on the DM mass, and result in much weaker exclusion as compared to the Dirac DM model. We also find that the combination of both vector and axial-vector couplings in the Dirac DM model leads to an interplay of constraints from direct detection and unitarity violation. Exclusion from strong spin-independent direct detection signals, given a vector coupling, overlaps with unitarity violation from the axial-vector coupling to give an exclusion that would not be present when fixing either to zero.

This paper is structured as follows. In Section \ref{sec:models}, we describe the different simplified DM models that we study. Section \ref{sec:constraints} contains a detailed description of the experimental likelihoods that we include in our analysis.  Section \ref{sec:results} contains the details of our global fit technique and our results, and Section~\ref{sec:conclusions} presents our conclusions. The samples from our scans and the corresponding \GB input files and plotting scripts can be downloaded from Zenodo \cite{Zenodo_DMsimp}.

\section{Models}
\label{sec:models}

We consider three models in this work, distinguished by the nature of the DM candidate:

\begin{itemize}
\item spin \(0\) (complex scalar)
\item spin \(\frac{1}{2}\) (Dirac)
\item spin \(\frac{1}{2}\) (Majorana)
\end{itemize}

In all cases, we assume that DM is a singlet under the SM gauge group. All three models have a spin \(1\) (vector) mediator. A thorough treatment of vector DM with a vector mediator requires the derivation of new unitarity limits, and we thus defer this option to Paper II of this series \cite{DMSimpII}. To enforce absolute stability of the DM candidate, we assume it to be odd under a new \(\mathbb{Z}_2\) symmetry under which the SM fields and the mediator are even. We take the mediator to be a real vector. This could typically arise in an extension of the SM gauge group by an abelian symmetry group such as U(1). We assume that none of the models gives rise to observable mixing between the mediator and any SM particles; this is required in order to reinterpret most of the experimental results that we consider. The models that we consider also assume that the mass generation mechanism that applies to the new particles has no observable effect on any of the experimental results that we consider. This could be achieved by, for example, the existence of a dark Higgs mechanism where the mass scale of the dark Higgs is well above both the mediator and DM masses. A detailed study of the case where this assumption breaks down for one model can be found in \cite{Duerr:2016tmh}. The models that we consider remain valid provided that the details of the encompassing higher order theory remain sufficiently decoupled at energy scales probed by current experiments.

For each model, we assume purely vector couplings to quarks (no axial-vector couplings) to prevent severe constraints from electroweak precision tests, and no lepton couplings\footnote{Leptophobic mediators could be accommodated in some extensions of the SM gauge group, such as gauged baryon number \cite{Duerr:2014wra} or anomaly-free extensions that require multiple new singlet fermions \cite{Ellis:2017tkh}.} to avoid restrictive dilepton searches \cite{Kahlhoefer:2015bea}. We also assume Minimal Flavour Violation to avoid constraints from flavour physics, and for simplicity, we assume equal couplings between up and down-type quarks. These together require universal quark couplings (\(g_\mathrm{q}\)) to the mediator.

\subsection{Scalar DM}

The Lagrangian density of our simplified model with a complex scalar DM candidate is
\begin{align}
\begin{split}
\mathcal{L}_\mathrm{BSM} =&  \partial_{\mu} \phi^{\dagger} \partial^{\mu} \phi - m_\mathrm{DM}^2 \phi^{\dagger} \phi \\
 - & \frac{1}{4} F^{\prime}_{\mu \nu} F^{\prime \mu \nu} - \frac{1}{2} m_\mathrm{M}^2 V_{\mu} V^{\mu}  + g_\mathrm{q} V_{\mu} \bar{q} \gamma^{\mu} q \\
 + & i g^{\mathrm{V}}_\mathrm{DM} V_{\mu} \Big(\phi^{\dagger} (\partial^{\mu} \phi) - (\partial^{\mu} \phi^{\dagger}) \phi \Big)\,,
\end{split}
\end{align}
where \(V_{\mu}\) is the mediator field, \(\phi\) is the scalar DM candidate and \(F^{\prime}_{\mu \nu}\) is the mediator field strength tensor. The model has four free parameters: the DM mass $m_\mathrm{DM}$, the mediator mass $m_{\mathrm{M}}$, the mediator-WIMP coupling $g_\mathrm{DM}^{\mathrm{V}}$ and the mediator-quark coupling $g_\mathrm{q}$. The DM candidate must be a complex scalar to avoid vanishing $g^{\mathrm{V}}_\mathrm{DM}$. This model is similar to one introduced in Ref.~\cite{Boehm:2003hm}, differing by the absence of lepton couplings.

The decay width of the mediator to quark $q$, for all three models is
\begin{align}
\label{eq:MediatorDecayWidth}
\Gamma(V_{\mu} \rightarrow \bar{q} q) = \frac{|g_\mathrm{q}|^2 m_\mathrm{M}}{4 \pi} \sqrt{1 - \frac{4 m_\mathrm{q}^2}{m_\mathrm{M}^2}} \Bigg( 1 + \frac{2 m_\mathrm{q}^2}{m_\mathrm{M}^2} \Bigg)\,,
\end{align}
while the decay width to the scalar DM candidate is
\begin{align}
\Gamma(V_{\mu} \rightarrow \phi^{\dagger} \phi) = \frac{|g_\mathrm{DM}^{\mathrm{V}}|^2 m_\mathrm{M}}{48 \pi} \Bigg(1 - \frac{4 m_\mathrm{DM}^2}{m_\mathrm{M}^2} \Bigg)^{3/2}.
\end{align}

\subsection{Dirac Fermion DM}

The simplified model with Dirac fermion DM has the Lagrangian density
\begin{align}
\begin{split}
\mathcal{L}_\mathrm{BSM} =& i \bar{\chi} \gamma^{\mu} \partial_{\mu} \chi - m_\mathrm{DM} \bar{\chi} \chi \\
- & \frac{1}{4} F^{\prime}_{\mu \nu} F^{\prime \mu \nu} - \frac{1}{2} m_\mathrm{M}^2 V_{\mu} V^{\mu} +   g_\mathrm{q} V_{\mu} \bar{q} \gamma^{\mu} q \\
+ & V_{\mu} \bar{\chi} (g^{\mathrm{V}}_\mathrm{DM} + g^{\mathrm{A}}_\mathrm{DM} \gamma^{5}) \gamma^{\mu} \chi \,,
\end{split}
\end{align}
where \(\chi\) represents the fermion DM candidate and the mediator and quark fields are given as before. The model has five free parameters, including the WIMP and mediator masses and the mediator-quark coupling, defined as before. Note, however, that this time there are two possible mediator-WIMP couplings, representing vector ($g^{\mathrm{V}}_\mathrm{DM}$) and axial-vector ($g^{\mathrm{A}}_\mathrm{DM}$) couplings. We vary \(g^{\mathrm{V}}_\mathrm{DM}\) and \(g^{\mathrm{A}}_\mathrm{DM}\) independently in our scans, allowing for possible interference effects between interactions mediated by the two.

The decay width to a given quark is given by Eq.\ \ref{eq:MediatorDecayWidth} while the decay width to the Dirac fermion DM candidate is
\begin{align}
\begin{split}
\Gamma(V_{\mu} & \rightarrow  \bar{\chi} \chi)  =  \frac{m_\mathrm{M}}{12 \pi} \sqrt{1 - \frac{4 m_\mathrm{DM}^2}{m_\mathrm{M}^2}} \\
& \times \Bigg(  |g^{\mathrm{V}}_\mathrm{DM}|^2 \Big( 1 + \frac{2 m_\mathrm{DM}^2}{m_\mathrm{M}^2} \Big) + |g^{\mathrm{A}}_\mathrm{DM}|^2 \Big(1 - \frac{4 m_\mathrm{DM}^2}{m_\mathrm{M}^2} \Big) \Bigg).
\end{split}
\end{align}

The axial-vector coupling in the fermion DM model implies that perturbative unitarity may be violated. The violation of unitarity stems from the omission of a dark Higgs boson in the simplified model and is analagous to the unitarity violation in the Standard Model without a Higgs boson. We adopt the bound of Ref.~\cite{Kahlhoefer:2015bea} for this:
\begin{align}
\label{SVFUnitarity}
m_\mathrm{DM} \leq \sqrt{\frac{\pi}{2}} \frac{m_\mathrm{M}}{g_\mathrm{DM}^{\mathrm{A}}}.
\end{align}
We apply this as an additional constraint, rejecting as theoretically invalid any parameter points that do not satisfy Eq.\ \ref{SVFUnitarity}.

\subsection{Majorana Fermion DM}

The model Lagrangian of the BSM additions  in the case of Majorana fermion DM is:

\begin{align}
\begin{split}
\mathcal{L}_\mathrm{BSM} =& \frac{1}{2} i \bar{\psi} \gamma^{\mu} \partial_{\mu} \psi - \frac{1}{2} m_\mathrm{DM} \bar{\psi} \psi \\
- & \frac{1}{4} F^{\prime}_{\mu \nu} F^{\prime \mu \nu} - \frac{1}{2} m_\mathrm{M}^2 V_{\mu} V^{\mu} +   g_\mathrm{q} V_{\mu} \bar{q} \gamma^{\mu} q \\
+ & \frac{1}{2} g^{\mathrm{A}}_\mathrm{DM} V_{\mu} \bar{\psi} \gamma^{5} \gamma^{\mu} \psi \,,
\end{split}
\end{align}

Unlike the Dirac Fermion case, Majorana DM prevents pure vector couplings to the mediator, and so this model has four free parameters. Like the Dirac fermion case, perturbative unitarity will be violated due to the presence of axial-vector couplings. The same bound (Eq.\ \ref{SVFUnitarity}) on parameters is applicable in this case, to exclude regions where unitarity is violated.

The decay width to a given quark is given by Eq.\ \ref{eq:MediatorDecayWidth} while the decay width to the Majorana fermion DM candidate is
\begin{align}
\Gamma(V_{\mu}  \rightarrow  \bar{\psi} \psi)  =  \frac{|g^{\mathrm{A}}_\mathrm{DM}|^2 m_\mathrm{M}}{12 \pi} \Bigg( 1 - \frac{4 m_\mathrm{DM}^2}{m_\mathrm{M}^2} \Bigg)^{3/2}.
\end{align}

\section{Constraints}
\label{sec:constraints}

DM-SM interactions are constrained by a broad range of astrophysical, cosmological and particle physics measurements.

We have implemented likelihoods for DM direct and indirect detection experiments, the measurement of the DM relic abundance and collider searches with the ATLAS and CMS experiments. For this we employ \gum \cite{GUM}, the GAMBIT Universal Model Machine. \gum generates the necessary model-specific module functions for \GB \textsf{2.3}, along with interfaces to the relevant backend codes that contain calculations for each DM observable. For the fermionic DM models, we supplement these likelihoods with the unitarity bound of Eq.\ \ref{SVFUnitarity}.

In Table~\ref{tab:experiments} we provide a summary of the likelihoods for different measurements sensitive to BSM physics included in our scans.  For searches where BSM effects are sought above some existing SM-induced event rate, we also give the value $\ln \mathcal {L}^\text{bg}$ that each likelihood takes in the pure SM case, where BSM physics is absent and only background events are observed.  In other cases, i.e.\ likelihoods corresponding to values of parameters of the SM itself, or of the relic density of DM, we simply give the best-case likelihood that results from predictions or parameters exactly matching their centrally measured values.

\begin{table}[t]
  \centering
  \begin{tabular}{lcc}
    \toprule
    \textbf{Experiment} & \textbf{$\ln \mathcal {L}^\text{bg}$} & \textbf{$\ln \mathcal {L}^\text{max}$}    \\ \midrule
        CDMSlite~\cite{Agnese:2015nto} & $-16.68$ \\
        CRESST-II~\cite{Angloher:2015ewa} & $-27.59$ \\
        CRESST-III~\cite{Abdelhameed:2019hmk} & $-27.22$ \\
        DarkSide 50~\cite{Agnes:2018fwg} & $-0.09$ \\
        LUX 2016~\cite{LUXrun2} & $-1.47$ \\
        PICO-60~\cite{Amole:2017dex,Amole:2019fdf} & $-1.496$ \\
        PandaX~\cite{Tan:2016zwf,Cui:2017nnn} & $-3.436$ \\
        XENON1T~\cite{Aprile:2018dbl} & $-3.651$ \\
        LZ 2022~\cite{LZ:2022ufs} & 0
        \\[2mm]
        LHC Dijets~\cite{CMS:2019gwf,ATLAS:2019fgd,ATLAS:2018qto,CDF:2008ieg,ATLAS:2018hbc,ATLAS:2018hzj,CMS:2019emo,ATLAS:2019itm,CMS:2019xai} & $0$ \\
        ATLAS monojet~\cite{Aad:2021egl} & $0$ \\
        CMS monojet~\cite{CMS:2021snz} & $0$
        \\[2mm]
        \emph{Fermi}-LAT~\cite{LATdwarfP8} & $-33.245$ \\
        \emph{Planck} 2018: $\Omega h^2$~\cite{Aghanim:2018eyx} & & $5.989$
        \\[2mm]
        Nuisances (see Table~\ref{tab:parameters}) & & $6.728$
        \\[1mm] \bottomrule
  \end{tabular}
  \caption{A summary of all likelihoods included in our scans.  For likelihoods that search for DM events above an SM background, we also give the SM-only (i.e.\ background-only) log-likelihood $\ln \mathcal {L}^\text{bg}$ for comparison.  For the remaining likelihoods, where there is no `SM background` to sensibly speak of, we give the maximum achievable value of the log-likelihood $\ln \mathcal {L}^\text{max}$, which corresponds to the case of a `perfect fit' where the predicted value of the relic density or the chosen value of a nuisance parameter is exactly equal to its measured value.}
  \label{tab:experiments}
\end{table}

The details for these likelihoods, and their implementation in \GB, are given in the following subsections.

\subsection{Relic Abundance}
\label{sec:Relic_Abundance}

The number density $n_{\scaleto{\mathrm{DM}}{3pt},\textrm{eq}}$ of DM particles in thermal equilibrium in the early Universe changes with time according to the Boltzmann equation~\cite{Gondolo:1990dk}
\begin{align} \label{eq:Boltzmann}
  \frac{dn_{\scaleto{\mathrm{DM}}{3pt}}}{dt} + 3Hn_{\scaleto{\mathrm{DM}}{3pt}} = -\langle\sigma v_\textrm{rel}\rangle \left(\left(n_{\scaleto{\mathrm{DM}}{3pt}} n_{\bar{\scaleto{\mathrm{DM}}{3pt}}}\right)-\left(n_{\scaleto{\mathrm{DM}}{3pt}}n_{\bar{\scaleto{\mathrm{DM}}{3pt}}}\right)_{\textrm{eq}}\right) \, ,
\end{align}
where $H(t)$ is the Hubble rate and \(n_{\scaleto{\mathrm{DM}}{3pt}} n_{\bar{\scaleto{\mathrm{DM}}{3pt}}} \equiv n_{\scaleto{\mathrm{DM}}{3pt}}^2\) for Majorana DM. $\langle\sigma v_{\textrm{{rel}}}\rangle$ is the thermally averaged cross-section times the relative velocity and is given by
\begin{align} \label{eq:sigmavthermal_def}
  \langle\sigma v_\textrm{rel}\rangle = \int^{\infty}_{4m_{\scaleto{\mathrm{DM}}{3pt}}^2} \!ds \, \frac{\sqrt{s-4m_{\scaleto{\mathrm{DM}}{3pt}}^2}(s-2m_{\scaleto{\mathrm{DM}}{3pt}}^2)\,K_1 \left(\frac{\sqrt{s}}{T}\right)}{8m_{\scaleto{\mathrm{DM}}{3pt}}^4 \,T K_2^2\left(\frac{m_{\scaleto{\mathrm{DM}}{3pt}}}{T}\right)} \,\sigma v_{\rm lab} \, ,
\end{align}
where $K_{1,2}$ are the modified Bessel functions and $v_{\rm lab}$ is the velocity of one of the annihilating
DM particles or antiparticles in the rest frame of the other (see Ref.~\cite{Binder:2017rgn} for further discussion). Dirac fermion WIMPs give a total contribution to the observed DM density of $n_{\scaleto{\mathrm{DM}}{3pt}}+n_{\bar{\scaleto{\mathrm{DM}}{3pt}}}=2n_{\scaleto{\mathrm{DM}}{3pt}}$ (where a possible initial asymmetry has been neglected~\cite{Kaplan:2009ag}).

We use \gum to generate tree-level annihilation cross-sections for each model using \CH \textsf{v3.6.27} \cite{Pukhov:2004ca,Belyaev:2012qa}. We obtain the WIMP relic density by using the \darkbit interface to \darksusy \textsf{v6.2.5}~\cite{darksusy,darksusy4} to numerically solve Eq.~\eqref{eq:Boltzmann} at each parameter point, assuming no departures from the standard cosmological history. For alternative scenarios, see Ref.~\cite{Arbey:2021gdg}. We compare the resulting value to the \emph{Planck} 2018 measurement of $\Omega_{\textrm{DM,obs}}\,h^2 = 0.120 \pm 0.001$~\cite{Aghanim:2018eyx}. We include a $1\%$ theoretical error on our computed relic density values, added in quadrature with the quoted \emph{Planck} uncertainty. More details on this prescription can be found in Refs.~\cite{gambit,DarkBit}.

Given the Planck measurement of the DM relic abundance, it is interesting to consider both the case where our hypothesised WIMPs constitute all of DM, and the case where they form only a subcomponent. In the former case, we use the Planck measurement to define a Gaussian likelihood based on the predicted WIMP abundance. In the latter case, we modify the likelihood such that it is flat if the predicted value is smaller than the observed one; details can be found in Ref.\ \cite{gambit}.

We rescale all predicted direct detection signals by the fraction
\begin{align}
f_\textrm{DM} \equiv  \Omega_\textrm{DM} / \Omega_{\mathrm{DM,obs}}.
\label{eq:fDM}
\end{align}
where \(\Omega_\textrm{DM} \equiv \Omega_{\phi} + \Omega_{\phi^{\dagger}} = 2 \Omega_{\phi} \) for scalar DM, \(\Omega_\textrm{DM} \equiv \Omega_{\chi} + \Omega_{\bar{\chi}} = 2 \Omega_{\chi}\) for Dirac DM and \(\Omega_\textrm{DM} \equiv \Omega_{\psi}\) for Majorana DM. Similarly, we rescale all indirect detection signals from DM-DM annihilation by $f_\mathrm{DM}^2$. Note that this encodes the two assumptions that $f_\mathrm{DM}$ is the same in all astrophysical systems, and the conservative assumption that any extra DM components are not visible to any of the astrophysical experiments that we consider.

\subsection{Direct Detection}

Direct detection experiments search for the scattering of DM particles off nuclei in some highly pure detector material, by measuring the nuclear recoil energy spectrum. The differential event rate with respect to the recoil energy $E_R$ is
\begin{align}
\frac{\text{d}R}{\text{d}E_\text{R}} = \frac{\rho_0}{m_\mathrm{T} \, m_{\mathrm{DM}}}  \int_{v_\text{min}}^\infty v f(v) \frac{\text{d} \sigma}{\mathrm{d} E_{\text{R}}}  \text{d}^3 v\; ,
\label{eq:dRdE}
\end{align}
where $\rho_0$ is the local density of DM at our point in the Galactic halo,
 $m_\mathrm{T}$ is mass of each target nucleus and $f(v)$ describes the local velocity distribution of the DM. Nuclear recoils of energy $E_R$ occur above some minimum DM velocity, which is given by
\begin{align}\label{eq:vmin}
v_\text{min}(E_\text{R}) = \sqrt{\frac{m_T E_\text{R}}{2 \, \mu^2}},
\end{align}
with $\mu = m_\mathrm{T} \, m_{\scaleto{\mathrm{DM}}{3pt}} / (m_\mathrm{T} + m_{\scaleto{\mathrm{DM}}{3pt}})$ the reduced mass of the DM-nucleus system.

Both the local density and the local velocity distribution of DM are subject to sizeable uncertainties. We therefore include the relevant quantities as nuisance parameters in our scans (see Section~\ref{sec:nuisances}). The remaining part of Eq.\ ~\ref{eq:dRdE} is the differential scattering cross-section $\mathrm{d}\sigma / \mathrm{d}E_\mathrm{R}$. In calculating this, one can make use of the knowledge that the scattering of WIMPs from the Galactic halo on nuclei is non-relativistic, typically involving momentum transfers below 200 MeV. If the mediator mass is much heavier than this scale (as assumed in our study), the mediator can be integrated out of the theory, and its contribution to scattering processes can instead be modelled using a non-relativistic EFT written as

\begin{align}
 \mathcal{L}_\text{NR} = \sum_{\mathrm{i},N} c_{\mathrm{i}}^\mathrm{N}(q^2) \, \mathcal{O}^\mathrm{N}_{\mathrm{i}} \; ,
 \label{eq:NREFT}
\end{align}
where the higher-dimensional operators $\mathcal{O}^\mathrm{N}_{\mathrm{i}}$ depend only on the spins of the WIMP and the nucleon ($\vec{S}_{\scaleto{\mathrm{DM}}{3pt}}$ and $\vec{S}_\mathrm{N}$), the momentum transfer $\vec{q}$ and the DM-nucleon relative velocity $\vec{v}$~\cite{Fitzpatrick:2012ix,Anand:2013yka,Fan:2010gt}. A systematic classification of the non-relativistic operators that can result from the reduction of simplified models can be found in Refs~\cite{Dent:2015zpa,Fitzpatrick:2012ix}, giving 16 operators in total.

One can then proceed by translating the parameters of the simplified model to the coefficients of the relevant operators $c_{\mathrm{i}}^{\mathrm{N}}(q^2)$ in the non-relativistic EFT. The relevant operators for our simplified models are provided in Table~\ref{DDoperators}. The associated prefactors are passed to \textsf{DDCalc v2.2.0}~\cite{DarkBit,HP} (through the $\ddm$ code, following Appendix A in \cite{GAMBIT:2021rlp}). We use \textsf{DDCalc} to compute the differential cross-section for each operator $\mathcal{O}^\mathrm{N}_{\mathrm{i}}$ and target element of interest, and to perform the velocity integral in Eq.\ \ref{eq:dRdE} in order to obtain the differential event rate. \ddcalc then implements detector effects and computes the final predicted number of events $N_\mathrm{p}$ at each experiment, by convolving the differential rate with the product $\phi(E_\mathrm{R})$ of the energy resolution and detector acceptance,
\begin{align}
 N_\mathrm{p} = M \, T_\text{exp} \int \phi(E_\mathrm{R}) \, \frac{\mathrm{d}R}{\mathrm{d}E_\mathrm{R}} \, \mathrm{d}E_\mathrm{R} \, ,
\end{align}
where $M$ is the detector mass and $T_\text{exp}$ is the exposure time.

In this work, we use direct detection data from the most recent XENON1T analysis~\cite{Aprile:2018dbl}, LUX 2016~\cite{LUXrun2}, PandaX 2016 and 2017~\cite{Tan:2016zwf,Cui:2017nnn}, CDMSlite~\cite{Agnese:2015nto}, CRESST-II and CRESST-III~\cite{Angloher:2015ewa, Abdelhameed:2019hmk}, PICO-60 2017 and 2019~\cite{Amole:2017dex,Amole:2019fdf}, DarkSide-50 \cite{Agnes:2018fwg} and LZ~\cite{LZ:2022ufs} \footnote{To implement the LZ likelihood, we take the publicly available efficiency function and consider only events below the mean of the nuclear recoil band, which effectively reduces the exposure by a factor of 2. We find  that the published LZ bound is accurately reproduced for DM masses above 100 GeV under the assumption that LZ observed no signal-like events. For smaller DM masses, our approach is unable to reproduce the downward fluctuation in the background, which leads to a stronger observed exclusion limit than the expected one. Our LZ limit is therefore conservative in this mass range, which is however inconsequential for our analysis.}.

\begin{table}
\centering
\begin{tabular}{ccc}
\toprule
Interaction              & Effective Operator                                                                                                    & Relevant models \\ \midrule
scalar                    & \(1_{\mathrm{DM}} 1_{\mathrm{N}}\) & Scalar \\ \hline
vector & \(1_{\mathrm{DM}} 1_{\mathrm{N}} \) & Dirac     \\ \hline 
axial-vector    & \(i \hat{\textbf{S}} \cdot \left(\hat{\textbf{S}}_{\mathrm{N}} \times \frac{\hat{q}}{m_{\mathrm{N}}} \right) \), \(\hat{\textbf{S}} \cdot \hat{\textbf{v}}^{\perp} 1_{\mathrm{N}}\) & Dirac, Majorana \\ 
\bottomrule \end{tabular}
\caption{Effective Operator matching to each model. The axial-vector couplings are momentum or velocity suppressed. Couplings to each operator follow \cite{Baum:2017kfa}.}
\label{DDoperators}
\end{table}

\subsection{Indirect Detection}

The processes that led to DM being in thermal equilibrium in the early Universe also allow for annihilation in the present day. Annihilation products can thus potentially be seen originating from regions of high DM density.  Of the possible products, gamma rays are particularly useful to search for, given that they should point back to the source without deflection. In recent years, both satellite and ground-based gamma ray observations have been used to constrain the possible interactions of DM.

An especially strong set of constraints on annihilating DM comes from observations of dwarf spheroidal galaxies~\cite{Bringmann:2012ez}. For a binned histogram of the DM-induced $\gamma$-ray flux, the flux from a target $k$ in bin $i$ can be written in the form $\Phi_i \cdot J_k$, where $\Phi_i$ includes the relevant particle physics, and $J_k$ includes the relevant astrophysics (see Ref~\cite{DarkBit} for more details).

The velocity dependence of the annihilation cross-section
\begin{align}
\langle \sigma v \rangle = a + b v^2 + \mathcal{O}(v^4),
\end{align}
plays a strong role in the sensitivity of indirect detection searches to different models.  If $a=0$, the leading-order term in the annihilation cross-section is proportional to \(v^2\) ($p$-wave annihilation), and the low average DM velocities in the dwarf spheroidal galaxies will result in a suppression of the predicted gamma-ray signals, compared to velocity-independent $s$-wave annihilation (where $a\ne0$). The two primary channels of annihilation are through the s-channel to a pair of quarks, and the t-channel to a pair of mediator particles.

In the case of scalar DM, annihilation to quarks is $p$-wave and annihilation to mediators is $s$-wave. When the latter channel is open ($m_\mathrm{DM} > m_\mathrm{M}$), it will dominate gamma-ray signals. The Dirac DM model has dominant $s$-wave annihilations to both quarks and mediators. Without vector couplings to DM or axial-vector quark couplings, the Majorana DM model's annihilation into quarks has no $s$-wave contribution. Like the scalar DM model, annihilation is $s$-wave to mediators when $m_\mathrm{DM} > m_\mathrm{M}$ \cite{Arcadi:2017kky}. Only $s$-wave contributions are large enough to impact searches towards dwarf spheroidals in the models that we consider, so we do not include the $p$-wave contributions in our gamma-ray flux predictions.  The Dirac fermion is therefore the only one of our models that leads to significant indirect detection signals when $m_\mathrm{DM} < m_\mathrm{M}$.

The $s$-wave contribution gives a particle physics factor of
\begin{align}
 \Phi_i &= \frac{f_\mathrm{DM}^2}{N_\mathrm{DM}} \sum_{j} \frac{(\sigma v)_{0,j}}{4\pi m_{\scaleto{\mathrm{DM}}{3pt}}^2}\int_{\Delta E_i} dE \, \frac{dN_{\gamma,j}}{dE} \, ,
\end{align}
where $(\sigma v)_{0,j}$ denotes the zero-velocity limit of the cross-section for $\mathrm{DM},\bar{\mathrm{DM}}\to j$, $N_{\gamma,j}$ is the number of photons that results from the final state channel $j$ (per annihilation), and $f_\mathrm{DM}$ is the DM fraction defined in Eq.\ ~\ref{eq:fDM}. The prefactor of \(1/N_\mathrm{DM}\) reflects the nature of the DM candidate: \(N_\mathrm{DM} = 2\) for self-conjugate particles and \(N_\mathrm{DM} = 4\) otherwise (assuming $n_{\scaleto{\mathrm{DM}}{3pt}}=n_{\bar{\scaleto{\mathrm{DM}}{3pt}}}$).

We use \CH to compute the annihilation cross-sections for each model, using the \gum interface. Each annihilation channel can produce either primary or secondary photons; the yields ${dN_{\gamma,j}}/{dE}$ are provided by \darkbit based on tabulated \pythia runs provided by \ds.

The astrophysics factor $J_k$ for each dwarf spheroidal galaxy $k$ is given by the line-of-sight integral over the DM distribution, assuming an NFW DM halo profile and the solid angle $\Omega$,

\begin{align}
  J_k &= \int_{\Delta\Omega_k} d\Omega \int_{\mathrm{l.o.s.}} ds \, \rho_{\rm DM}^2\simeq
  D_k^{-2} \int d^3x\,\rho_{\rm DM}^2\,, \label{eq:def_jk}
\end{align}
where $D_k$ is the distance to the galaxy. We use the \texttt{Pass-8} combined analysis of 15 dwarf spheroidal galaxies performed by the \emph{Fermi}-LAT Collaboration using 6 years of
LAT data~\cite{LATdwarfP8}, computing the likelihood with the \darkbit interface to \gamlike~\textsf{v1.0.1}. The log-likelihood $\ln \mathcal{L}_{\rm{exp}}$ is constructed from the product $\Phi_i \cdot J_k$ summed over all targets and energy bins,
\begin{align}
 \ln \mathcal{L}_{\rm{exp}} = \sum^{\rm{N_{dSphs}}}_{k=1} \sum^{\rm{N_{eBins}}}_{i=1} \ln \mathcal{L}_{ki}\left(\Phi_i \cdot J_k\right) \, .
\end{align}
An additional likelihood contribution comes from treating the $J_k$ factors of each dwarf spheroidal galaxy as nuisance parameters, giving $\ln \mathcal{L}_J = \sum_k \ln \mathcal{L}(J_k)$~\cite{DarkBit,LATdwarfP8}. The full likelihood, profiled over the J factors, is then given by
\begin{align}
 \ln \mathcal{L}_{\rm{dSphs}}^{\rm{prof.}} = \underset{\{J_k\}}{\textrm{max}}\left(\ln\mathcal{L}_{\rm{exp}} + \ln\mathcal{L}_J \right) \, .
\end{align}

An alternative to dwarf spheroidal measurements is to look for evidence of DM annihilation in the centre of our Galaxy. Although \emph{Fermi}-LAT Galactic Centre limits are not nearly as robust as those from dwarf spheroidal galaxies, the forthcoming Cherenkov Telescope Array (CTA) is expected to probe thermally-produced WIMPs up to particle masses of several TeV~\cite{Acharyya:2020sbj}. We briefly consider the future impact of CTA observations on the viable parameter space of our simplified models in Section \ref{section:future_prospects}.

\subsection{Collider searches for WIMPs using monojet events}

The simplified models defined in Section~\ref{sec:models} allow for pair production of WIMPs in proton-proton collisions at the LHC. This process becomes visible if one of the incoming partons radiates a jet through initial state radiation, giving a potential signal of events with a single jet plus missing transverse energy ($\slashed{E}_{\mathrm{T}}$). In this study, we include CMS and ATLAS searches for monojet events based respectively on $137\,\rm{fb}^{-1}$ \cite{CMS:2021snz} and $139\,\rm{fb}^{-1}$ \cite{Aad:2021egl} of Run II data. We neglect other signatures such as monophoton events, which are known to give weaker constraints on our simplified models than monojet searches~\cite{Bauer:2017fsw,Zhou:2013fla,Brennan:2016xjh}.

For a given parameter point of a simplified model, the key theory input to the monojet search likelihood for an LHC experiment is the set of predicted event yields in each bin of the missing transverse energy distribution. In each bin, the yield is given by
\begin{align}
N = L\times\sigma \times(\epsilon A)\;,
\end{align}
where $L$ is the integrated luminosity, $\sigma$ is the total production cross-section, and $\epsilon A$ is the product of the efficiency and acceptance for passing the kinematic selections that define the analysis.

The quantity $\epsilon A$ can be obtained by combining a Monte Carlo simulation of DM production with a simulation of the ATLAS/CMS detector. The standard approach to this in \GB is to run the \pythia Monte Carlo generator at each point in the global fit to simulate events, followed by a fast detector simulation based on four-vector smearing with typical resolution functions \cite{ColliderBit}. The problem for monojet events, however, is that the expected signatures crucially depend on the ISR model used to simulated jet radiation in order to correctly predict the transverse missing energy distribution~\cite{Buckley:2014fba}. We have therefore performed a more detailed simulation using \textsf{MadGraph\_aMC@NLO}~\cite{Alwall:2011uj} (\textsf{v3.1.1}), interfaced to \pythia \textsf{v8.3} \cite{Sjostrand:2007gs} for parton showering and hadronization. We use the CKKW prescription to perform the matching between \MG and \pythia. We computed matrix elements for \MG starting from Universal FeynRules Output (UFO)  files \cite{Degrande:2011ua}, generated with \fr \cite{Alloul:2013bka} and employing a 5-flavour scheme. We used \madanalysis 5 \cite{Conte:2012fm} to perform detector simulation and implement each of the ATLAS and CMS monojet analyses in order to compute $\epsilon A$. As this set of simulations is too computationally expensive to run during the global fit, we precomputed grids of the cross sections (\(\sigma\)) and $\epsilon A$ factors for each LHC experiment in advance, and interpolated them at runtime using \colliderbit in order to obtain predicted LHC signal yields.  We then fed the predicted yields to the likelihood functions contained in \colliderbit in order to obtain the final constraints.

Our interpolation grids were defined as follows:
\begin{itemize}
\item \textbf{mediator mass}: 20 values, 50\,GeV--10\,TeV
\item \textbf{DM/mediator mass ratio}: 31 values, 0.1--40
\item \textbf{quark-mediator coupling}: 6 values, 0.01--1.0
\item \textbf{DM-mediator couplings}: 7 values (each), 0.01--3.0
\end{itemize}

We used the ratio of DM and mediator masses, rather than the masses themselves, so as to be able to include a higher density of points across the resonance region, where rapid changes in signal prediction occur. In order to avoid simulating points unnecessarily, we did not simulate parameter combinations that violate Eq.\ \ref{SVFUnitarity}. We limit the DM mass/mediator mass ratio to 0.1, as below this one can safely extrapolate to small DM masses. After imposing the unitarity requirement on the grid points, the total numbers of points are 26040 (Scalar DM), 8880 (Majorana DM) and 62160 (Dirac DM).

The CMS analysis that we include has 66 exclusive signal regions. The Collaboration have published a covariance matrix that allows all of these to be used simultaneously in constructing the likelihood function. We use the ``simplified likelihood'' method \cite{Collaboration:2242860}, which approximates the full experimental likelihood function by a Poisson counting term convolved with a multi-dimensional Gaussian likelihood describing the correlated systematic uncertainties on the background predictions:
\begin{align}
  \label{eq:simplike}
  \begin{split}
    \mathcal{L}_{\text{CMS}}(\bm{s}, \bm{\gamma})
    =& \prod_{i=1}^{66} \left[ \frac{(s_i + b_i + \gamma_i)^{n_i} \, e^{-(s_i + b_i + \gamma_i)}}{n_i!} \right]\\
    & \hphantom{\int} \times \frac{1}{\sqrt{\det2\pi\Sigma}} e^{-\frac{1}{2} \bm{\gamma}^T \bm{\Sigma^{-1}} \bm{\gamma}} \, .
  \end{split}
\end{align}
For each signal region $i$, the observed counts, expected signal and expected background yields are represented by $n_i$, $s_i$ and $b_i$, respectively. The term $\gamma_i$ quantifies the deviation from the nominal expected yields due to systematic uncertainties in signal region $i$.  The set $\{\gamma_i\}$ thus gives 66 nuisance parameters in total for this particular analysis. The covariance matrix $\bm{\Sigma}$ provided by CMS encodes the correlations between the various $\gamma_i$ factors.  We supplement these by adding the signal yield uncertainties in quadrature along the diagonal. For every parameter point, we profile out the 66 $\gamma_i$ parameters so that the final CMS likelihood is defined solely in terms of the simplified model signal estimates $\bm{s}$:
\begin{align}
  \mathcal{L}_{\text{CMS}}(\bm{s}) \equiv \mathcal{L}_{\text{CMS}}(\bm{s}, \hat{\hat{\bm{\gamma}}}),
\end{align}
where $\hat{\hat{\bm{\gamma}}}$ denotes the combination of background nuisance parameters resulting in the highest value of the likelihood for the given signal $\bm{s}$.

The ATLAS analysis does not come with a published covariance matrix, nor with a published likelihood in the HistFactory format of Ref.\ \cite{ATL-PHYS-PUB-2019-029}. The conservative course of action is therefore to calculate a likelihood using the signal region with the best expected sensitivity. To maximize the sensitivity of this procedure, we combine the three highest missing energy bins so that $\slashed{E}_{\mathrm{T}} > 1000 \, \mathrm{GeV}$ is the highest bin in the analysis. This is justified by the fact that the systematic uncertainties in the background estimations for these bins (and hence their correlations) are negligible. The ATLAS likelihood is then given by
\begin{align}
  \mathcal{L}_{\text{ATLAS}}(s_i) \equiv \mathcal{L}_{\text{ATLAS}}(s_i, \hat{\hat{\gamma_i}}) \, ,
\end{align}
where $\mathcal{L}_{\text{ATLAS}}(s_i, \hat{\hat{\gamma_i}})$ is the single-bin equivalent of Eq.~(\ref{eq:simplike}), whilst $i$ labels the signal region that would give the lowest likelihood in the case $n_i = b_i$.

Assuming that the ATLAS and CMS searches are independent, we then calculate a total log-likelihood for LHC monojet searches as $\ln \mathcal{L}_{\text{LHC}} = \ln \mathcal{L}_{\text{CMS}} + \ln \mathcal{L}_{\text{ATLAS}}$.

Note, however, that the choice of different ATLAS signal regions for different parameter points leads to a large variation in the effective likelihood normalisation between different parameter points. The standard \colliderbit solution is to instead set the LHC log-likelihood contribution to the total difference in log-likelihood between the signal and background-only ($\bm{s} = \bm{0}$) cases:
\begin{align}
  \Delta \ln \mathcal{L}_{\text{LHC}} = \ln \mathcal{L}_{\text{LHC}}(\bm{s}) - \ln \mathcal{L}_{\text{LHC}}(\bm{s}=\bm{0}).
  \label{eq:LHC_loglike}
\end{align}
Positive values of this quantity indicate that a DM model fits the observed data better than the background-only hypothesis. In cases where there are small excesses in the LHC data, this can lead to regions of the simplified model parameter space fitting the data better than other regions that might be indistinguishable from the SM.  As our global fit results are presented as $1\sigma$ and $2\sigma$ confidence regions defined using the likelihood ratio $\mathcal{L}(\bf{\Theta}) / \mathcal{L}(\bf{\Theta}_\text{best-fit})$ around the best-fit parameters $\bf{\Theta}_\text{best-fit}$, this can exclude parameter regions that exhibit only a little worse agreement with the data than the SM. Whilst this is of course the correct result in the case where one takes excesses at face value and attempts to fit them, it can also be instructive to consider LHC results under the conservative assumption that any excesses simply arise from statistical fluctuations rather than BSM physics. We therefore run separate scans for this latter case, where we ``cap'' the LHC likelihood as
\begin{align}
  \label{eq:LHC_loglike_capped}
  \Delta \ln \mathcal{L}_{\text{LHC}}^\text{cap}(\bm{s}) &=
  \min[\Delta \ln \mathcal{L}_{\text{LHC}}({\bm s}),
       \Delta \ln \mathcal{L}_{\text{LHC}}({\bm s}={\bm 0})]\nonumber\\
       &= \min[\Delta \ln \mathcal{L}_{\text{LHC}}({\bm s}),0].
\end{align}
More detail on this procedure can be found in Ref.~\cite{EWMSSM}.

\subsection{Collider searches for the mediator using dijet events}

Because our simplified models explicitly include a mediator particle that interacts with SM quarks, it is possible for the LHC to produce a mediator that decays to quarks, rather than WIMPs. This should generate an excess of dijet events, each with a dijet invariant mass approximately equal to the mass of the mediator. Searches for dijet resonances thus provide powerful constraints on DM simplified models, though analyses have to employ various clever tricks to increase sensitivity given the extremely large multijet background at the LHC.

Assuming that a narrow width approximation holds, $\sigma (\mathrm{pp} \rightarrow V_{\mu} \rightarrow \mathrm{qq}) \approx \sigma (\mathrm{pp} \rightarrow V_{\mu}) \times BR(V_{\mu} \rightarrow \mathrm{qq})$. In this case, $\sigma \propto g_\mathrm{q}^2$, so we implement ATLAS and CMS dijet limits by appropriately scaling the published limits by the mediator-quark coupling and the branching ratio into quarks, following the same approach as Ref.~\cite{Bagnaschi:2019djj}. We interpolate these published limits in $m_\mathrm{M}$ during our scans and select the most constraining search for a given mediator mass. This way, we are able to recast the published limits without using Monte Carlo simulation.

\begin{figure}
\includegraphics[width=\columnwidth]{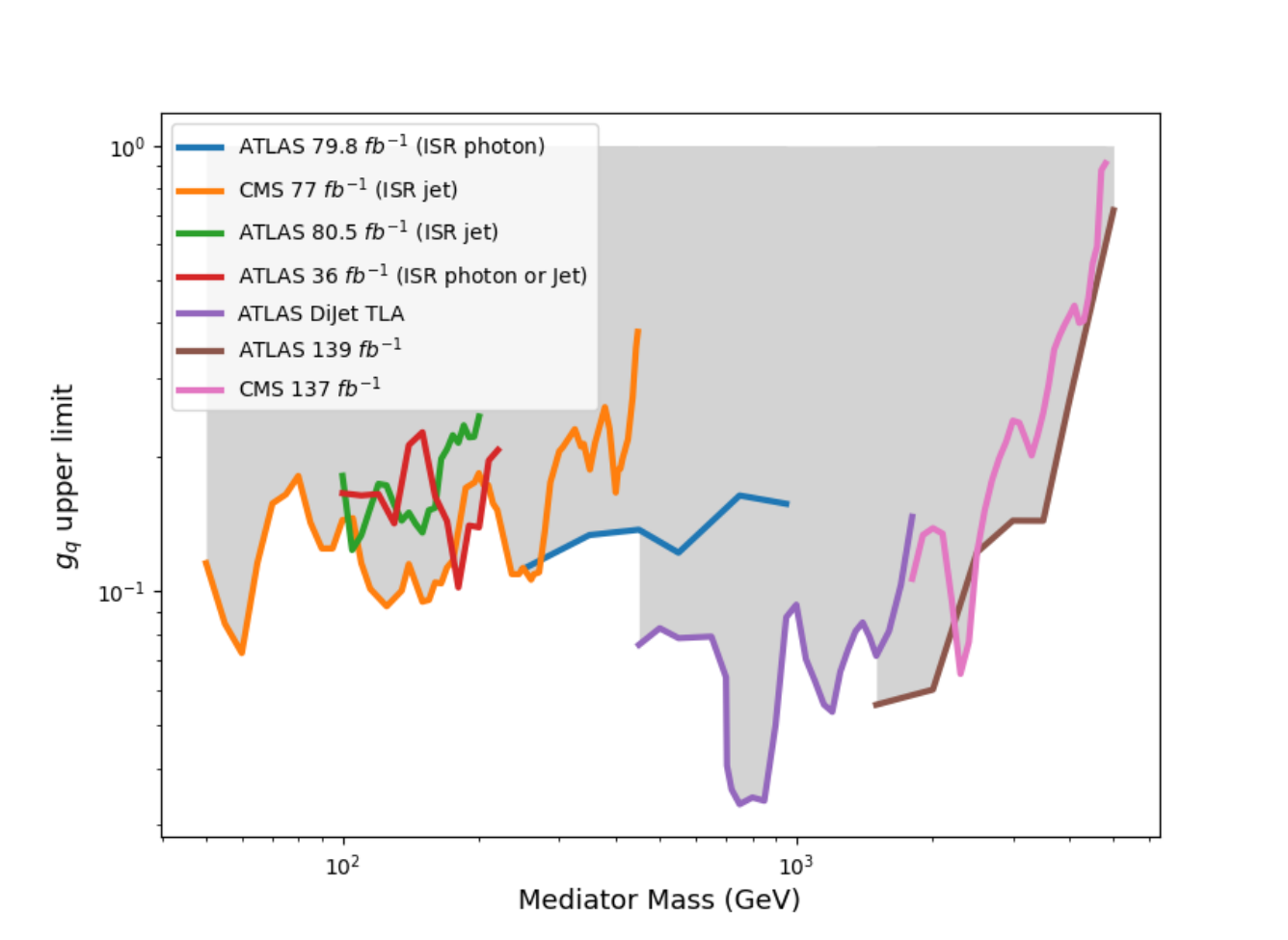}
\caption{Quark coupling upper limits from each dijet search included in our likelihood Eq.\ \ref{eq:lnlike_dijet}. For a given mediator mass, the \(95\%\) confidence dijet limit that we use in our likelihood is the one that is the most constraining (i.e. closest to the bottom of this plot).}
\label{fig:dijets}
\end{figure}

We then compare our results to the coupling upper limits ($g_\mathrm{q,excl}$) provided by a broad range of LHC dijet searches \cite{CMS:2019gwf,ATLAS:2019fgd,ATLAS:2018qto,CDF:2008ieg,ATLAS:2018hbc,ATLAS:2018hzj,CMS:2019emo,ATLAS:2019itm,CMS:2019xai}, using a likelihood of the form
\begin{align}
\ln{\mathcal{L}_\mathrm{dijet}} = -2 \Big( \frac{g_\mathrm{q,pred}^4 \times BR(V_{\mu} \rightarrow \mathrm{qq})^2}{g_\mathrm{q,excl}^4} \Big)\,.
\label{eq:lnlike_dijet}
\end{align}
The factor of \(-2\) arises because \(g_\mathrm{q,excl}\) is taken from the \(95\%\) confidence limit on the coupling provided by each analysis, corresponding to \(\Delta \chi^2 = 4\). This produces coupling upper limits as shown in Figure~\ref{fig:dijets}.

There is a high degree of variation with mediator mass in the limits presented from each analysis, due to the signal prediction moving in and out of different experimental analysis bins. The effect of these variations will not show up strongly in our later results, as we profile over the model couplings and the highest likelihood points will tend to be where dijet constraints are unconstraining.

\subsection{Nuisance parameter likelihoods}
\label{sec:nuisances}
The model parameters for each of our simplified models are supplemented by a series of nuisance parameters, which contribute to our astrophysical likelihoods. We give a complete list of nuisance parameters in Table~\ref{tab:parameters}.

Our treatment of the local DM density $\rho_0$ follows the default prescription in \darkbit, which assumes that $\rho_0$ is log-normally distributed with central value $\rho_0 = 0.40$\,GeV\,cm$^{-3}$ and error $\sigma_{\rho_0}=0.15$\,GeV\,cm$^{-3}$. The asymmetric scan range for $\rho_0$ in Table~\ref{tab:parameters} reflects this log-normal distribution. All other nuisance parameters are scanned over their $3 \sigma$ range as provided in Table~\ref{tab:parameters}.

We treat the Milky Way halo in the same way as in our previous studies of Higgs portal and DM EFTs~\cite{HP,DMEFT}. The DM velocity follows a Maxwell-Boltzmann distribution, with uncertainties on the peak of the distribution and the Galactic escape velocity described by Gaussian likelihoods with $v_{\rm{peak}} = 240\, \pm \,8$\,km\,s$^{-1}$ \cite{Reid:2014boa} and $v_{\rm{esc}} = 528 \pm 25$\,km\,s$^{-1}$ (based on \emph{Gaia} data \cite{Deason:2019kgj}), respectively.

\section{Results}
\label{sec:results}

We have performed comprehensive scans of each simplified model parameter space using the differential evolution sampler \textsf{Diver v1.0.4} \cite{ScannerBit} with a population of \(10\,000\) and a convergence threshold of \(10^{-6}\). We present results as profile likelihood maps in planes of the parameters and/or derived quantities. We carried out scans for the cases where the observed DM relic density is taken as an upper limit or as a two-sided measurement, and also for the cases where the LHC likelihood is capped or uncapped. This gives four combinations of scans for each model. The parameter points with the highest likelihoods for each model are given in Table \ref{tab:BestFitPoints:SVS}.

Table~\ref{tab:parameters} outlines our complete list of parameters and their associated scan ranges. We do not consider models with large fine-tuning or large hierarchies between the different couplings, as this may be challenging to achieve when considering plausible UV embeddings of the simplified models. In order to equally explore each order of magnitude in the Lagrangian parameters, we sampled them using log-uniform priors. For nuisance parameters, we sampled using flat priors. We stress, however, that these `prior' distributions play no formal role in the final statistical analysis of the profile likelihood maps that we present, but merely provide efficiently distributed starting guesses from which to hunt for best-fit points and map likelihood functions.

\begin{table*}
\centering
\begin{tabular}{ll}
\toprule
\textbf{Parameters}          & \textbf{Range} \\ \midrule
DM mass, \(m_\mathrm{DM}\)          & \([50,10000]$\,GeV     \\ \hline
Mediator mass, \(m_\mathrm{M}\)     & \([50,10000]$\,GeV    \\ \hline
quark-mediator coupling, \(g_\mathrm{q}\)     & \([0.01,1.0]\)    \\ \hline
mediator-DM coupling (vector), \(g^{\rm V}_\mathrm{DM}\)   & \([0.01,3.0]\) \\ \hline
mediator-DM coupling (axial vector), \(g^{\rm A}_\mathrm{DM}\) & \([0.01,3.0]\) \\ \midrule
\textbf{Nuisance Parameters} & \textbf{Value} ($\pm 3 \sigma$ range) \\ \midrule
Local DM density, \(\rho_0\)   & \([0.2,0.8]$\,GeV\,cm$^{-3}\)  \\ \hline
Most probable speed, \(v_\mathrm{peak}\) & \(240 (24)$\,km\,s$^{-1}\)     \\ \hline
Galactic escape speed, \(v_\mathrm{esc}\)   & \(528 (75)$\,km\,s$^{-1}\) \\
\bottomrule
\end{tabular}
\caption{Ranges scanned over for model and nuisance parameters. The axial-vector coupling is present only in the Dirac fermion model. Hadronic input parameters are given at \(\mu = 2$\,GeV.}
\label{tab:parameters}
\end{table*}

\begin{table*}[t]
\centering
\begin{tabular}{llcccccc}
\toprule
LHC & Relic Density & Best Fit \(m_\mathrm{DM}\) (GeV) & Best Fit \(m_\mathrm{M}\) (GeV) & Best Fit \(g_\mathrm{q}\) & Best Fit \(g^{\rm V}_\mathrm{DM}\) & Best Fit \(g^{\rm A}_\mathrm{DM}\)    & \(\Delta \ln \mathcal{L}\) \\ \midrule
 \multicolumn{2}{l}{\textbf{Scalar DM}}  & \\ \midrule
Full &  Upper limit   & 4965 & 10000 & 0.0100 & 2.333  & --  & -0.019  \\ \hline
Full &  All DM  & 4532 & 9203 & 0.0101 & 0.825 & -- & -0.469 \\   \midrule
\multicolumn{2}{l}{\textbf{Dirac fermion DM}} & \\ \midrule
Full & Upper limit & 262 & 537 & 0.0301 & 0.0100 & 0.990 & 4.48 \\  \hline
Capped & Upper limit & 146 & 300 & 0.0108 & 0.0103 & 2.525 & -0.089 \\  \hline
Full & All DM & 588 & 1320 & 0.0369 & 0.0100 & 0.754 & 0.881  \\  \hline
Capped & All DM & 3762 & 7744 & 0.0151 & 0.0118 & 0.536 & -0.559 \\ \midrule
\multicolumn{2}{l}{\textbf{Majorana fermion DM}} & \\ \midrule
Full & Upper limit & 50 & 114 & 0.0130 & -- & 0.243  & 4.779      \\  \hline
Capped & Upper limit & 668 & 1400 & 0.0139 & -- & 0.763  & 0.001 \\  \hline
Full & All DM & 69.1 & 166 & 0.0104 & -- & 0.307  & 3.12 \\  \hline
Capped & All DM & 2423 & 5079 & 0.0279 & -- & 0.125 & -0.449 \\  \bottomrule
\end{tabular}
\caption{Approximate best-fit points for each model. $\Delta \ln \mathcal{L}$ values are defined as $\ln \mathcal{L} - \ln \mathcal{L}^{\mathrm{bg}}$, where the background-only likelihood is detailed in Table \ref{tab:experiments}.}
\label{tab:BestFitPoints:SVS}
\end{table*}

\subsection{Scalar DM}

\begin{figure}
         \centering
         \includegraphics[width=\columnwidth]{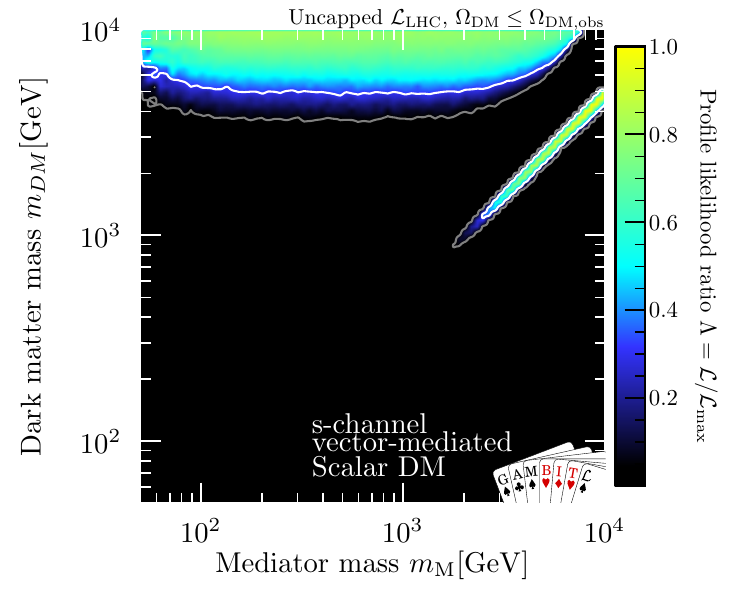}
         \includegraphics[width=\columnwidth]{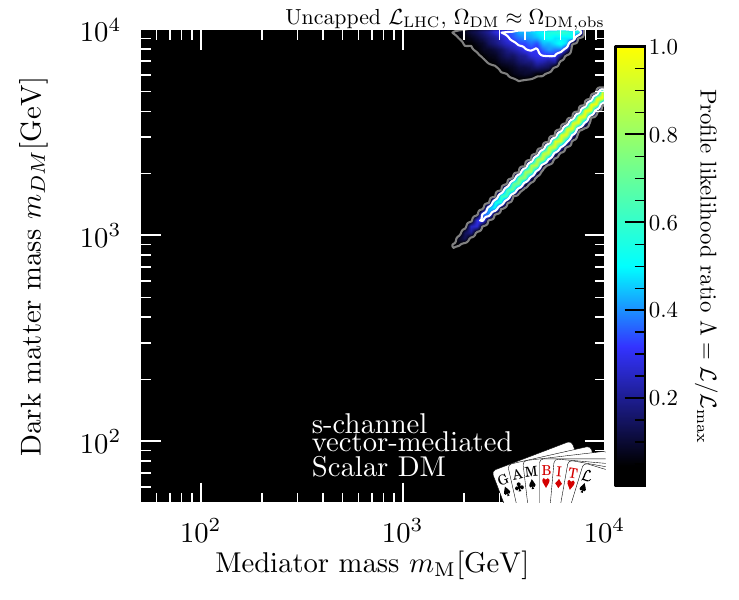}
         \caption{Scalar DM profile likelihood, profiling over couplings.  The observed relic density of DM is taken as an upper limit (top) or to consist entirely of the scalar DM candidate (bottom).  \(1\sigma\) and \(2 \sigma\) contours are shown in white and grey respectively.}
         \label{SVS:Combined_RD_exact}
\end{figure}

\begin{figure*}[tbp]
  \centering
  \includegraphics[width=\columnwidth]{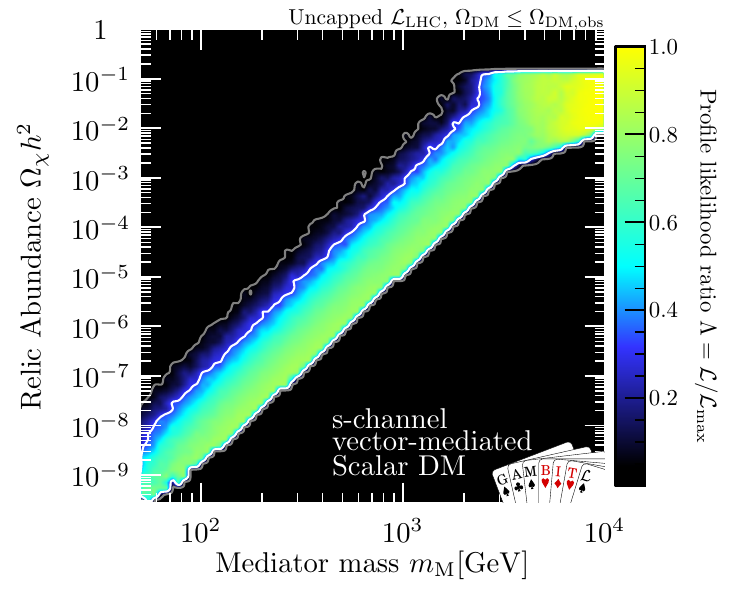}
  \includegraphics[width=\columnwidth]{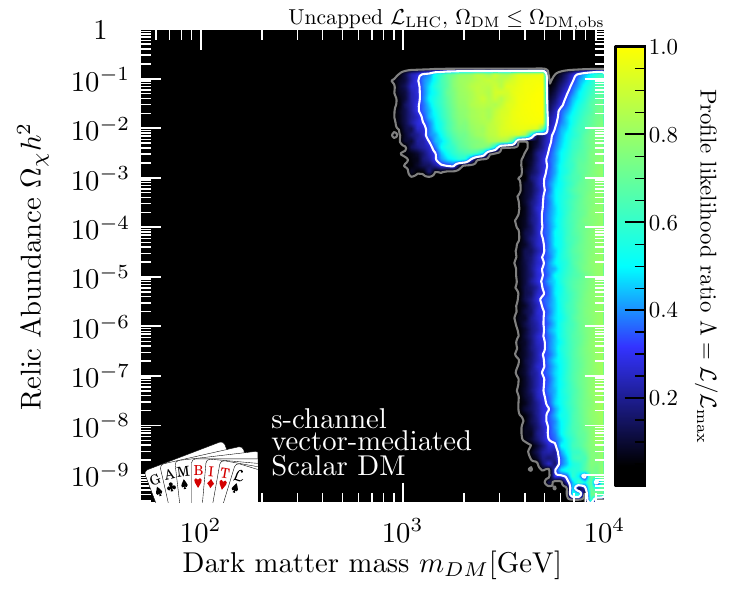}
  \caption{Relic Density of the scalar DM model as a function of the mediator mass (left) and DM mass (right). The results are shown for the scan where the observed relic density of DM is allowed to exceed the abundance of the scalar DM candidate. \(1\sigma\) and \(2 \sigma\) contours are shown in white and grey respectively.}
  \label{SVS:RelicDensity}
\end{figure*}

Results from global scans of the scalar DM model are shown in Figure~\ref{SVS:Combined_RD_exact}.  Any excesses present in the mono-jet likelihoods can only be fitted by models that are already robustly excluded by other searches, so we do not show any results for this model based on the capped collider likelihood.

The results show two separate regions allowed at the \(2\sigma\) level. The shape of the surviving parameter space is defined primarily by the relic abundance. The viable parameter space is split in two corresponding to the two DM annihilation channels, along the resonance for s-channel annihlation into a pair of quarks, and $m_\mathrm{DM} > m_\mathrm{M}$, where $t$-channel annihilation into a pair of mediators is possible. In regions either side of the diagonal resonance region, DM is overproduced and the model is inconsistent with relic density measurements. Highest likelihood points lie along the resonance, where \(2 m_\mathrm{DM} \approx m_\mathrm{M}\) and annihilation is enhanced, bringing the DM relic density down to or below the observed value.  This region is most preferred toward the upper mass limits of the scan, where the best-fit point lies (see Table~\ref{tab:BestFitPoints:SVS}).

When requiring that the scalar DM candidate explains all of DM (Figure~\ref{SVS:Combined_RD_exact}), mediator masses up to approximately \(2\)\,TeV are excluded. Toward lower mediator masses the strength of the effective coupling used to calculate direct detection signals is increased. To escape direct detection bounds, the highest likelihood moves toward regions that underproduce the relic abundance. This can be seen clearly in Figure~\ref{SVS:RelicDensity}, where we plot the relic density as a function of the mediator mass (left) and DM mass (right), in the scan where the model was allowed to underproduce the DM abundance.  Whether the relic density is taken as an upper limit or a two-sided constraint, DM masses below approximately \(1\)\,TeV are excluded. This exclusion is, of course, dependent on the limits of the quark coupling \(g_{\rm q}\) in the scan, and reducing this lower bound will expand the allowed region.

Direct detection limits give the lower bound on DM masses, along the resonance region in both scans and also when \(m_\mathrm{DM} > m_\mathrm{M}\) in Figure~\ref{SVS:Combined_RD_exact} (top) since these experiments are more constraining for lighter DM masses (provided that the mass is well above the energy threshold of the experiment). These experiments also drive the best-fit point towards the border of the scan, where the predicted signal decreases. Near the boundaries of the scan, this likelihood estimate is close enough to the zero signal likelihood that the magnitude of the profile likelihood ratios would not be noticeably changed by extending the scan limits to higher mediator masses.

Indirect detection limits give a slight preference to the region along the resonance. This is because when \(m_\mathrm{DM} > m_\mathrm{M}\), $t$-channel annihilation of DM to mediator particles (and subsequent decays to SM products) would produce an observable effect on gamma-ray searches. This signal would be in weak tension with the absence of a positive detection thus far. This effect is reduced toward the lower mediator masses, as gamma ray predictions are scaled by the relic abundance which significantly underpredicts the DM abundance for lower $m_\mathrm{M}$.

Dijet searches contribute by giving preference to models with lower quark couplings, but monojet searches do not have a strong influence on the overall profile likelihood at all.

Given that the likelihood is weakly dependent on the couplings, and dependent primarily on the ratio of the mediator and DM masses, a reasonable estimate for the number of effective degrees of freedom would be 1 or 2. We compute an approximate $p$-value of the best-fit likelihood against the background only scenario described in Table \ref{tab:experiments} whenever the background only scenario is preferred. At the best-fit point, for 1--2 effective degrees of freedom, the $p$-value is between $0.85$ and $0.98$ when allowing the scalar DM to underpredict the relic abundance, and between $0.33$ and $0.63$ when saturating the abundance. Neither of these are statistically distinct from the Standard Model.

\subsection{Dirac Fermion DM}

\begin{figure*}[t]
         \centering
           \includegraphics[width=\columnwidth]{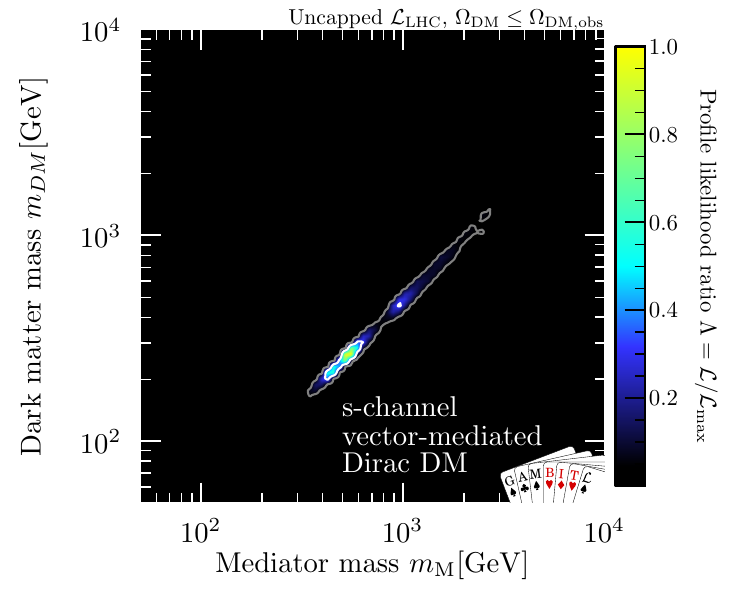}
           \includegraphics[width=\columnwidth]{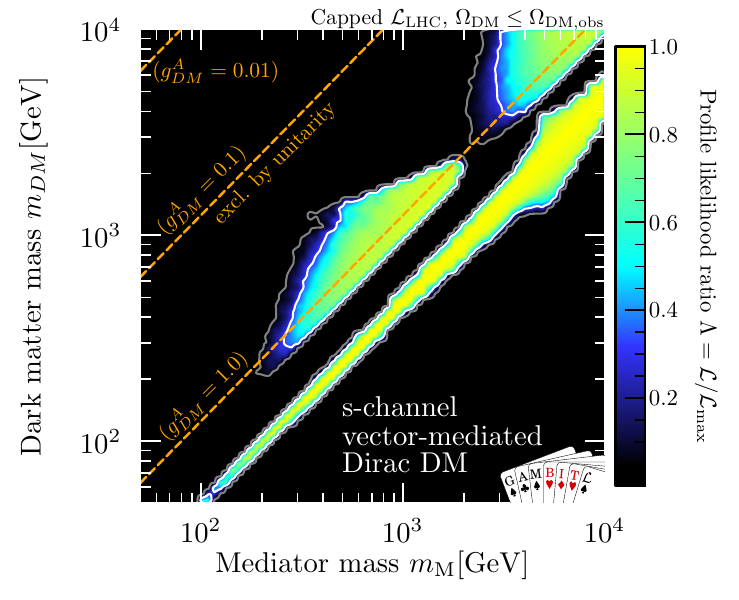}
           \includegraphics[width=\columnwidth]{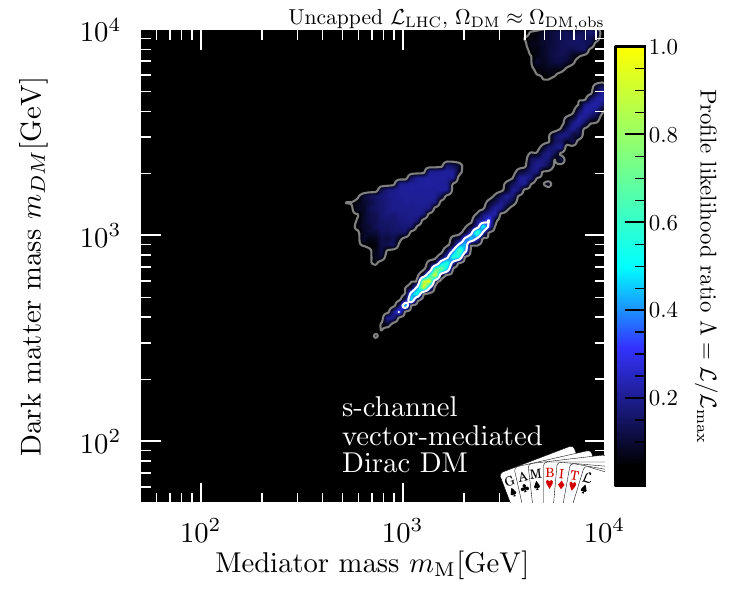}
           \includegraphics[width=\columnwidth]{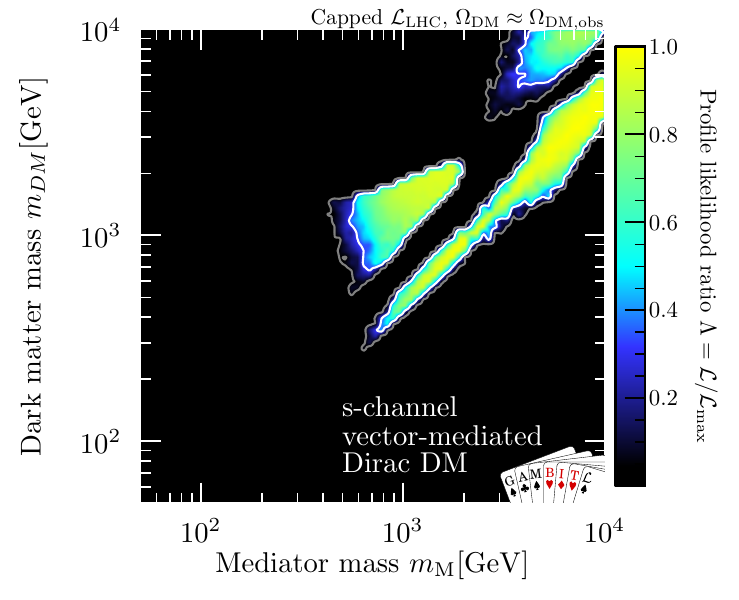}
           \caption{Profile likelihood for the Dirac fermion DM model. The observed relic density of DM is taken as an upper limit (top) or to consist entirely of the Dirac DM candidate (bottom). The collider likelihood is either uncapped (left), or capped to prevent preference over the Standard Model (right). \(1\sigma\) and \(2 \sigma\) contours are shown in white and grey respectively.}
           \label{SVF:ProfileLike}
\end{figure*}

The constraints on the Dirac fermion DM model are shown in Figure~\ref{SVF:ProfileLike}. As with the scalar DM, a large portion of the non-excluded parameter space lies on the resonance region where \(2 m_\mathrm{DM} \approx m_\mathrm{M}\).

\begin{figure*}[t]
\centering
\includegraphics[width=1.5\columnwidth]{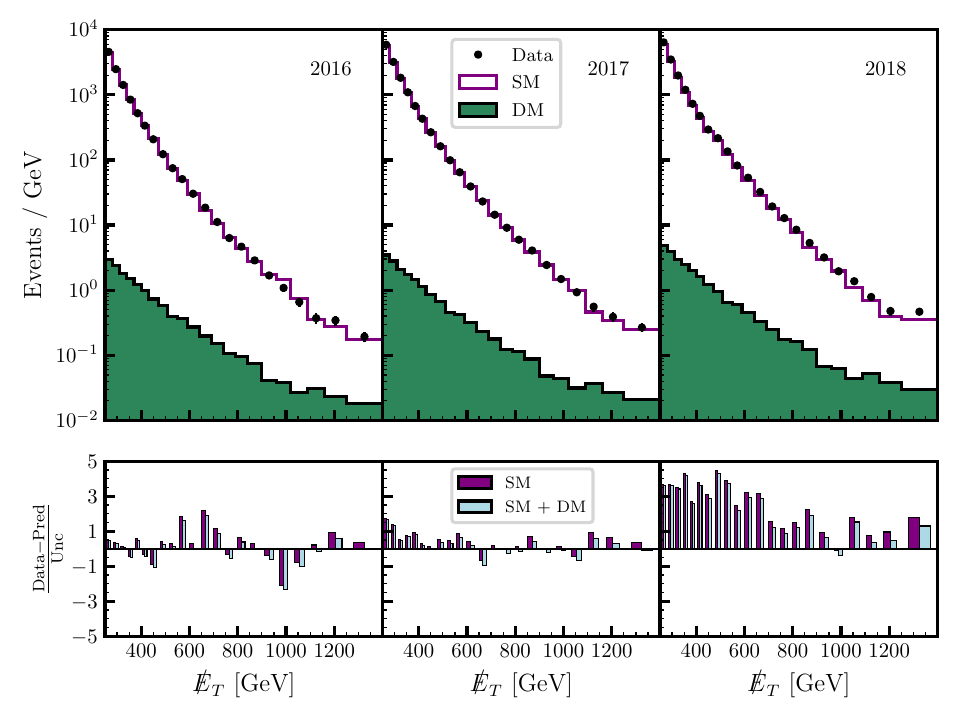}
\caption{Missing energy spectra for the CMS monojet search. The SM background prediction (purple) and the observed event counts (black) are taken from Ref~\cite{CMS:2021snz}. The green distribution shows the signal prediction for the best-fit Dirac DM model. The bottom panel shows the residuals, defined as (data - prediction)/uncertainty for both the SM and the SM + DM predictions.}
\label{SVF:METSpectra}
\end{figure*}

Excesses in the monojet collider searches are partially fit by the model. This leaves a highly-preferred region around a mediator mass of \(500\)\,GeV (Figure~\ref{SVF:ProfileLike}, top left). In Figure~\ref{SVF:METSpectra}, we show the signal for the best-fit point that would contribute to the CMS monojet search \cite{CMS:2021snz}. The signal regions in this search are split into three years of data taking, with a strong difference between SM prediction and data in the last year. This difference is present in the simplified likelihood only, but was not present in their joint fit of control and signal regions. The model cannot entirely fit the 2018 excess without strongly over predicting the signal in the two other years. If these excesses are assumed to simply be statistical fluctuations, and the monojet likelihood is capped (Figure~\ref{SVF:ProfileLike}, right), then the surviving parameter space opens up.

For the sake of illustration, we also show the bounds from unitarity that apply to this model in the top right panel of Figure~\ref{SVF:ProfileLike}. There is no strong preference toward the unitarity bound as the profile likelihood in this region is in fact very flat in $g^{\rm A}_\mathrm{DM}$ --- a fact reflected in the range of best-fit couplings shown in Table \ref{tab:BestFitPoints:SVS}.

\begin{figure*}[t]
  \centering
  \includegraphics[width=\columnwidth]{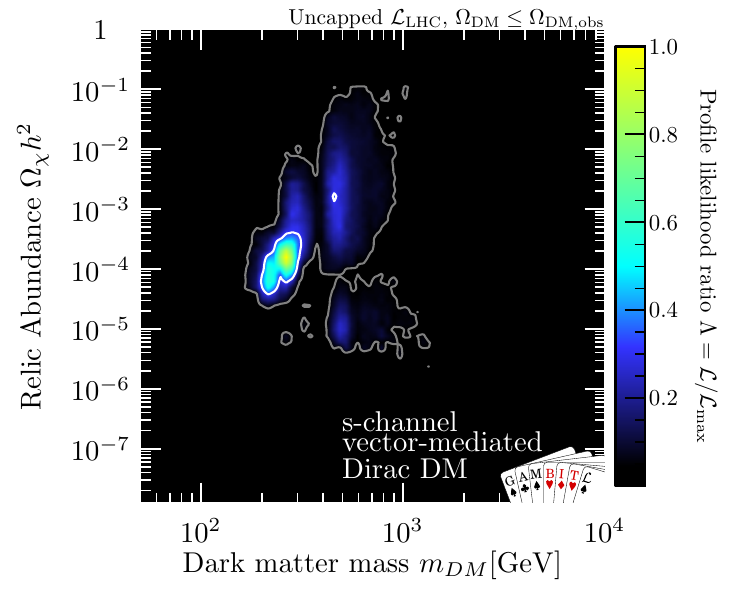}
  \includegraphics[width=\columnwidth]{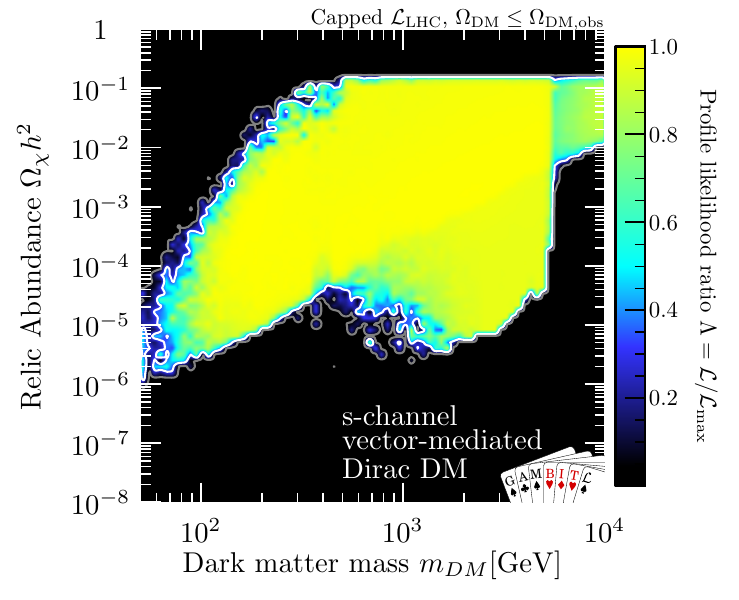}
  \vspace{-2mm}
  \caption{Profile likelihood as a function of relic density of the Dirac fermion DM model, allowing the model to underproduce the relic abundance. The capped collider result is shown on the right. We do not show the dependence on mediator mass as it does not differ greatly from the dependence on DM mass. \(1\sigma\) and \(2 \sigma\) contours are shown in white and grey respectively.}
  \label{SVF:RelicDensity}
\end{figure*}

Relic density limits have a strong influence on the exclusion contours, such that the allowed regions lie either on the resonance or, in the case where collider likelihoods are capped, where \(m_\mathrm{DM} > m_\mathrm{M}\). The model overproduces DM below the resonance region, where \(m_\mathrm{M} > 2 m_\mathrm{DM}\). Requiring that the DM candidate constitute all of the observed DM excludes regions along the resonance where $m_\mathrm{DM} \lesssim 300$ GeV and regions off the resonance where $m_\mathrm{M} \lesssim 600$ GeV, where the model cannot escape direct detection bounds without underproducing the DM abundance. This shifts the best-fit point to higher masses around DM mass of \(1300\) GeV, and reduces the ability of the best fit to fit the monojet excesses. This suppression of the likelihood of the best fit (compared to that found when imposing the relic density as an upper limit) opens up some of the parameter space off resonance at the very limits of the 2$\sigma$ region. For the results where we cap the collider likelihood, the best-fit likelihood is also reduced by requiring the DM abundance to be saturated, which broadens the surviving parameter space along the resonance region. Figure~\ref{SVF:RelicDensity} shows that if the monojet excesses were explained by this model, rather than by statistical fluctuations in the experiments, this particle would not saturate the DM relic abundance. Additional DM candidates would be required to explain the observed relic abundance.

The combination of direct detection, relic abundance and unitarity constraints provide the shape of the off-resonance region seen in the capped collider results (Figure~\ref{SVF:ProfileLike} right) and the uncapped collider results when requiring a saturated relic abundance (Figure~\ref{SVF:ProfileLike} bottom left). To avoid unitarity violation, the model is excluded for larger $g^{\rm A}_\mathrm{DM}$. This in turn may prevent sufficient annihilation and cause the predicted relic abundance to exceed the observed value. Direct detection experiments provide a lower bound on the mediator mass for a given DM mass and $g^{\rm V}_\mathrm{DM}$. Since the strength of the direct detection signals is primarily from the $g^{\rm V}_\mathrm{DM}$ coupling, and the unitarity bound is from the $g^{\rm A}_\mathrm{DM}$ coupling, this shape in parameter space would differ in either the pure vector or pure axial-vector coupling cases (as studied in Ref.~\cite{Bagnaschi:2019djj}). If the $g_{\rm q}$ limit was lowered, this allowed region would expand such that the two off-resonance regions would become one. Indirect detection searches do not play a strong role in the overall exclusion contours for this model.

Since the model is preferred over the Standard Model when the collider likelihood is uncapped, we only calculate a $p$-value for the capped collider scans. At the best-fit point, for 1--2 effective degrees of freedom, the $p$-value is between $0.67$ and $0.91$ when allowing the Dirac fermion to underpredict the relic abundance, and between $0.29$ and $0.57$ when saturating the abundance. These are not statistically distinct from the Standard Model.

\begin{figure}[tbp]
  \centering
  \includegraphics[width=\columnwidth]{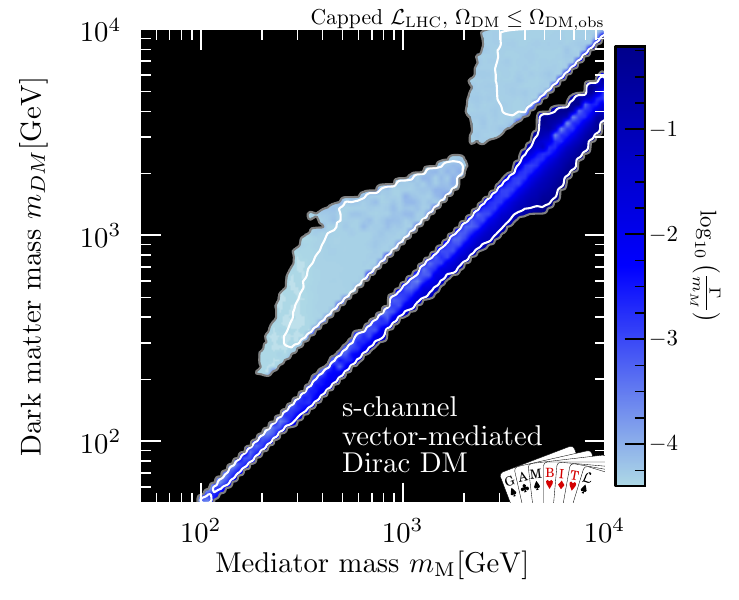}
  \vspace{-2mm}
  \caption{Ratio of mediator decay width to mass for all regions allowed in Figure~\ref{SVF:ProfileLike} (top right).}
  \label{SVF:NWA}
\end{figure}

For the narrow width approximation to hold, the ratio of mediator decay width to mediator mass must remain low. Figure~\ref{SVF:NWA} shows that this ratio increases with higher mediator masses and lower DM masses. In the surviving parameter space of the scan, this ratio can reach at most roughly 0.4--0.5. As this is close to the limit that would prevent accurate recasting of dijet search limits, doubt could be raised about the validity of applying these limits. However, this occurs in regions where dijet limits are unconstraining. In all regions where collider limits contribute noticeably, the model safely satisfies the narrow width approximation.

The results differ from those in Ref.~\cite{Bagnaschi:2019djj} as we present combined fits of all 5 model parameters varying concurrently, whereas they separate the model into pure vector/axial-vector cases. We also allow the model to fit monojet excesses and give an overall preference over the Standard Model, where they do not. In this way, this study is complementary to \cite{Bagnaschi:2019djj} without presenting duplication of their results.

\subsection{Majorana Fermion DM}

\begin{figure*}[t]
         \centering
           \includegraphics[width=\columnwidth]{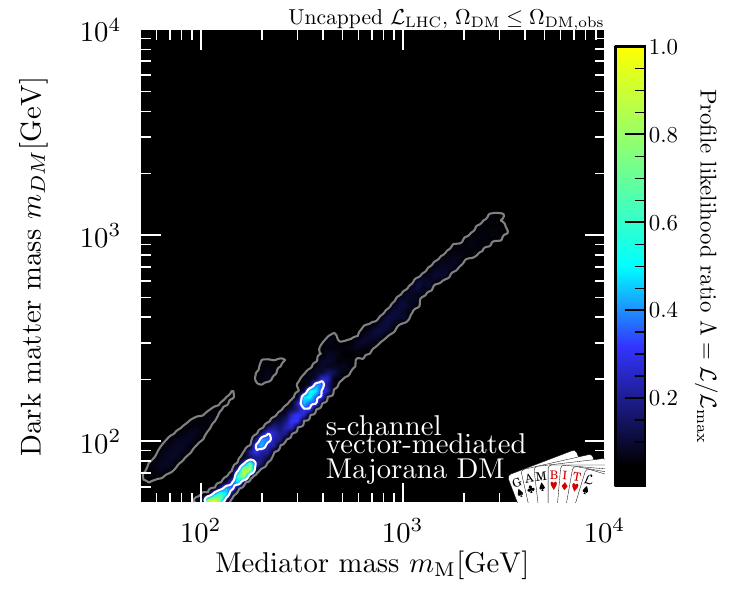}
           \includegraphics[width=\columnwidth]{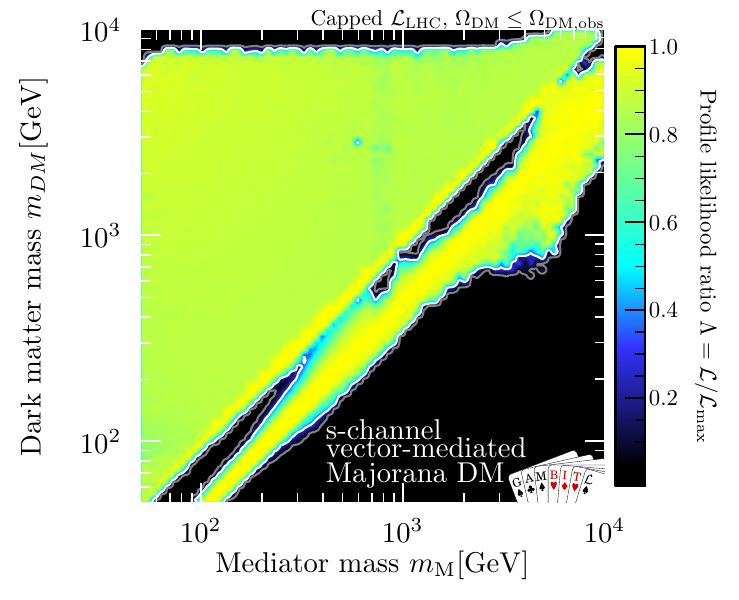}
           \includegraphics[width=\columnwidth]{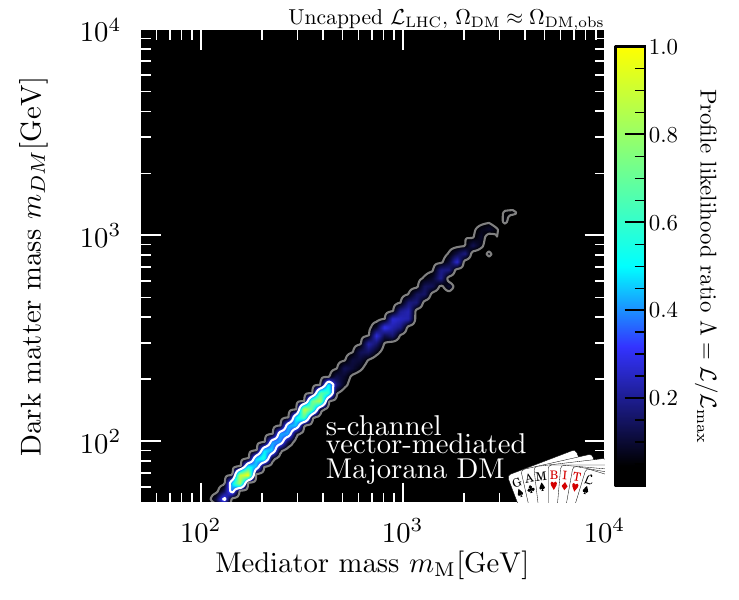}
           \includegraphics[width=\columnwidth]{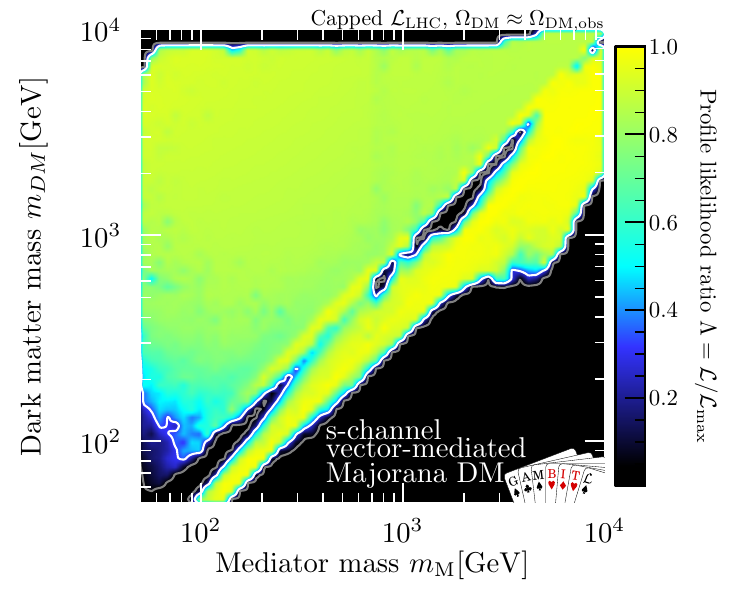}
           \caption{Profile likelihood for the Majorana fermion DM model. The observed relic density of DM is taken as an upper limit (top) or a two-sided measurement that the model must match (bottom). The collider likelihood is either uncapped (left), or capped to prevent preference over the Standard Model (right). \(1\sigma\) and \(2 \sigma\) contours are shown in white and grey respectively.}
           \label{SVMF:ProfileLike}
\end{figure*}

\begin{figure*}[t]
         \centering
         \includegraphics[width=\columnwidth]{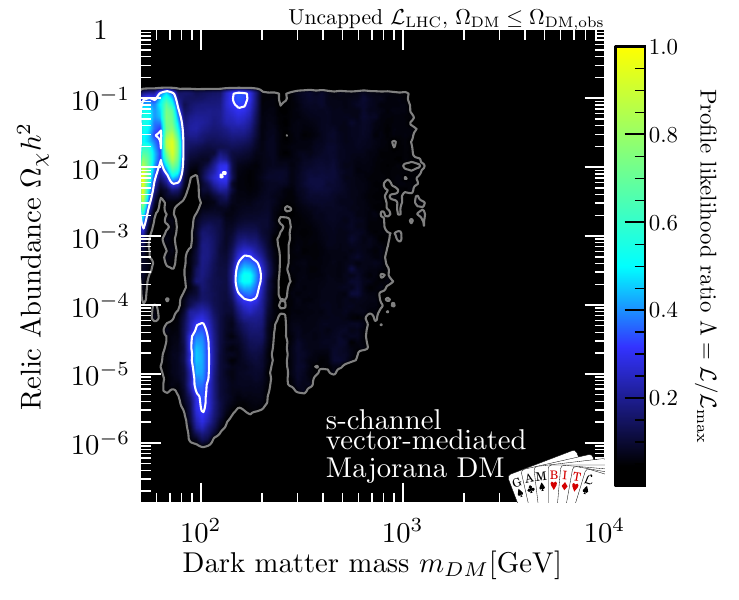}
         \includegraphics[width=\columnwidth]{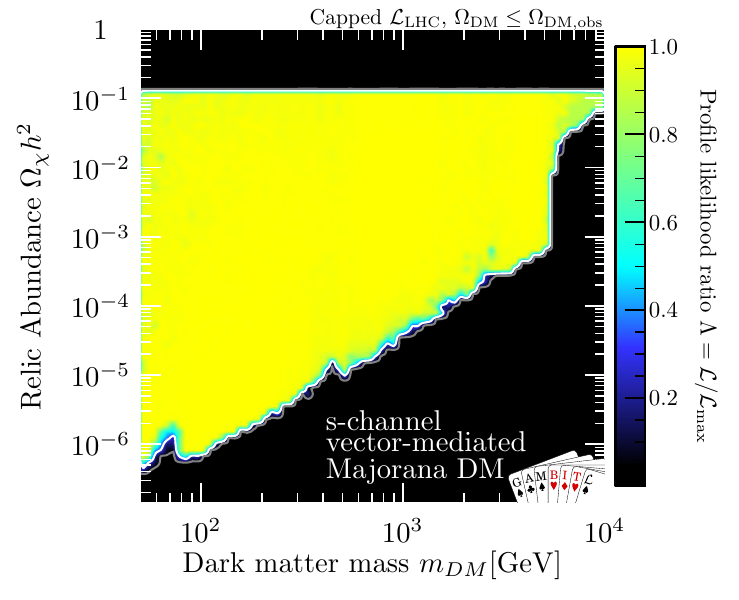}
         \caption{Relic density of the Majorana fermion DM model, allowing the model to underproduce the relic abundance. Capped collider results are shown on the right. As with the Dirac model, we do not show the dependence on mediator mass as it does not differ greatly from the dependence on DM mass. \(1\sigma\) and \(2 \sigma\) contours are shown in white and grey respectively.}
         \label{SVMF:RD}
\end{figure*}

Results from global scans of the Majorana fermion DM model are shown in Figure~\ref{SVMF:ProfileLike}. Like the Dirac fermion model, there is a strong preference over the background along the resonance. The collider excess is predominantly fit along the resonance region.

Figure~\ref{SVMF:ProfileLike} (right) shows how capping the collider likelihood expands the \(2 \sigma\) contours, with little exclusion when $m_\mathrm{DM} > m_\mathrm{M}$. The reason this region is much larger in the Majorana models compared to the Dirac model, is that the direct detection signals from axial-vector couplings are suppressed in the non-relativistic limit. The presence of vector couplings in the Dirac model give rise to strong enough direct detection signals such that this region is not allowed for the scanned coupling ranges, i.e.\ even with $g^{\rm V}_\mathrm{DM}=0.01$.

As with the Dirac fermion and scalar models, relic density limits play a strong role in determining the shape of the allowed regions such that abundance measurements exclude regions with a high mediator mass and low DM mass. Figure~\ref{SVMF:RD} shows the relic abundance as a function of the DM mass. The best-fit point appears to under-predict the relic abundance, although there is little preference for the best-fit point over a region that saturates the abundance. Requiring that the DM candidate saturates the relic abundance, in Figure~\ref{SVMF:ProfileLike} (bottom), shifts the location of the best-fit point slightly towards higher mediator/DM masses, however the capped collider results are very similar regardless of whether the abundance is saturated or not. The approximate $p$-value for 1--2 effective degrees of freedom is between $0.34$ and $0.64$ in the case of a saturated relic abundance and a capped collider likelihood. The best-fit points for the other three scans gave preferences over the Standard Model.

The preference for the resonance regions over the \(m_\mathrm{DM} > m_\mathrm{M}\) region in the fits with capped $\mathcal{L}_{\text{LHC}}$ is because in the latter region of parameter space, the gamma-ray flux is not negligible when the annihilation channel into mediators is open. This increase in the annihilation cross-section at late times is not matched by a drop in the relic density fraction, as $p$-wave annihilation dominates at early times. The result is that the \emph{Fermi}-LAT likelihood gives a preference to regions where \(m_\mathrm{DM} < m_\mathrm{M}\).

\subsection{Future prospects}
\label{section:future_prospects}

\begin{figure}
         \centering
         \includegraphics[width=\columnwidth]{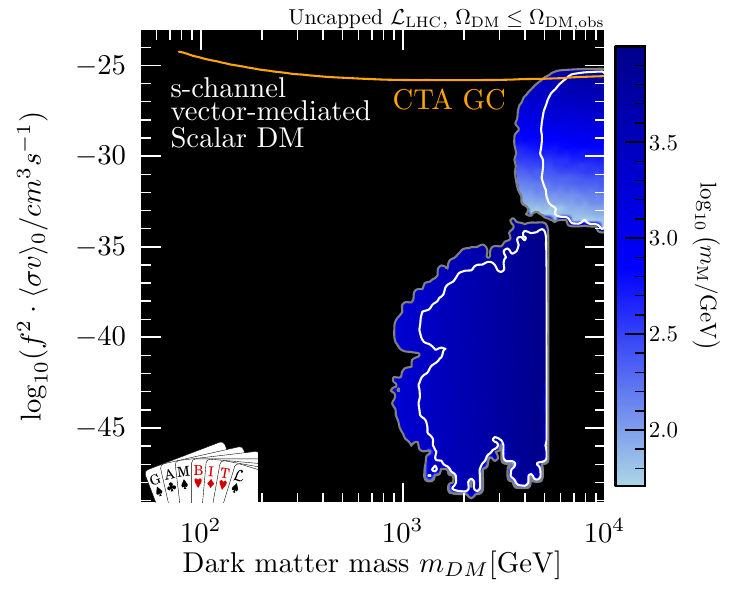}
         \includegraphics[width=\columnwidth]{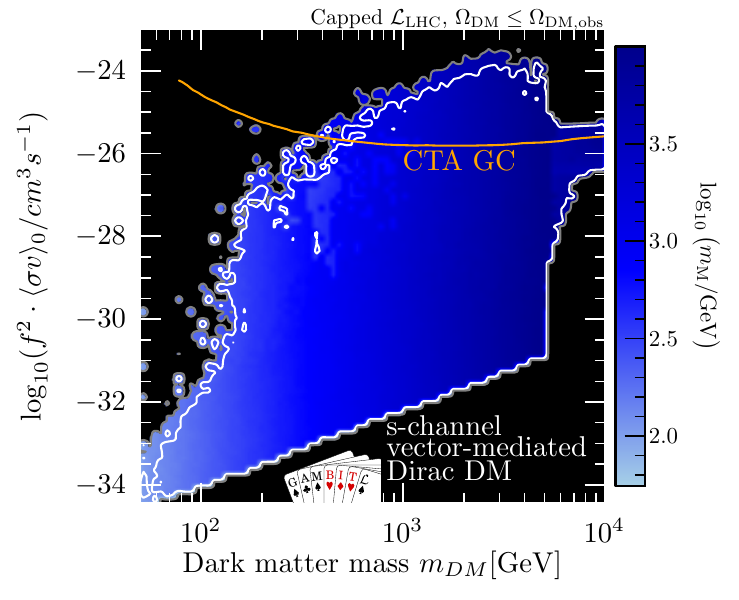}
         \includegraphics[width=\columnwidth]{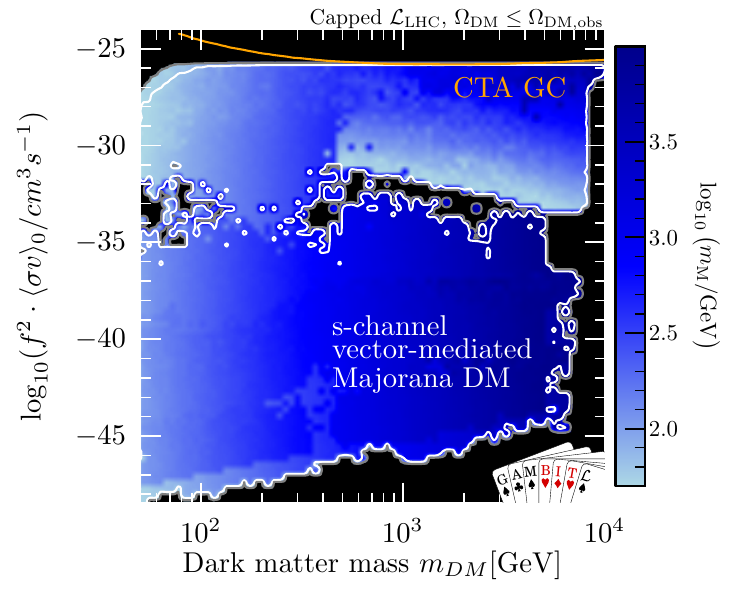}
         \caption{Allowed parameter space as a function of the DM mass and rescaled annihilation cross section. Capped collider likelihoods are shown for Dirac (middle) and Majorana (bottom) DM as these are the least constraining. The non-capped likelihood is shown for the scalar DM model (top). The solid red line indicates the projected ``initial construction'' sensitivity of the Cherenkov Telescope Array (CTA) towards the Galactic Centre (GC). The contour is coloured to indicate the mediator mass that gives the highest likelihood for a given DM mass. }
         \label{future_prospects:CTA}
\end{figure}

\begin{figure}
         \centering
         \includegraphics[width=\columnwidth]{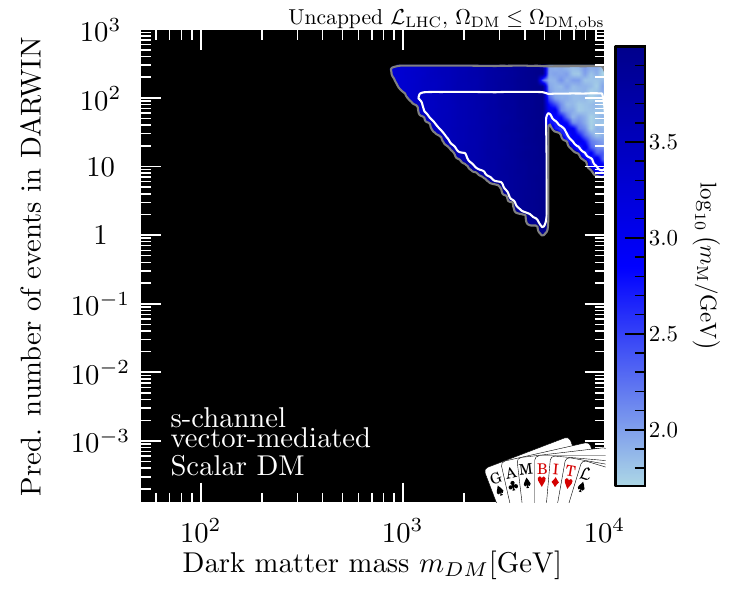}
         \includegraphics[width=\columnwidth]{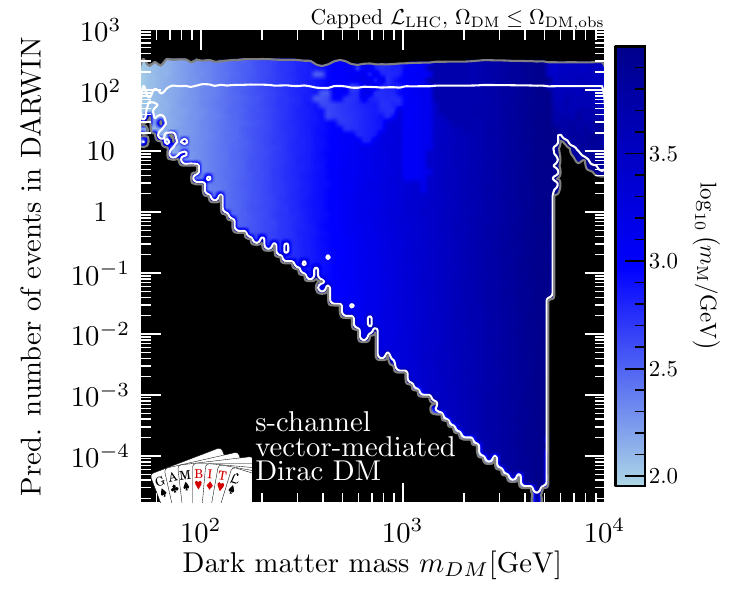}
         \includegraphics[width=\columnwidth]{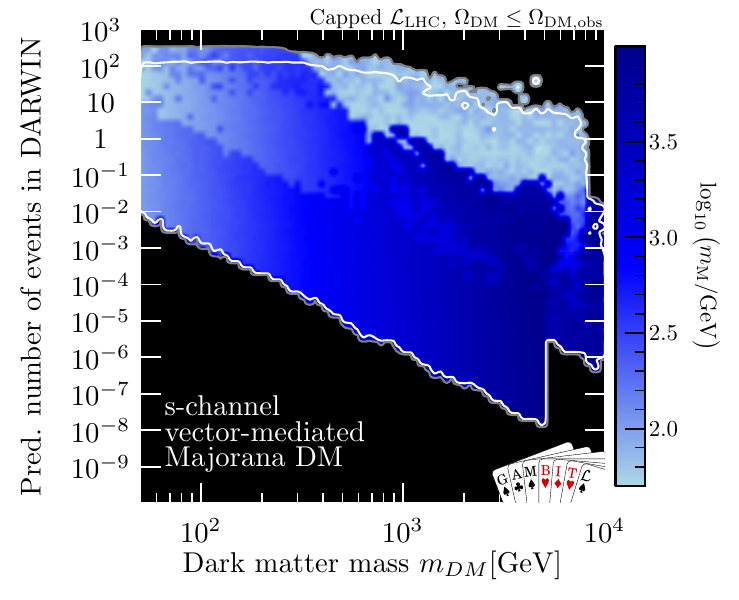}
         \caption{Mediator mass as a function of the predicted number of signal events in the DARWIN experiment. Capped collider likelihoods are shown for Dirac (middle) and Majorana (bottom) DM as these are the least constraining. The non-capped likelihood is shown for the scalar DM model (top). The contour is coloured to indicate the mediator mass that gives the highest likelihood for a given DM mass.}
         \label{future_prospects:DARWIN}
\end{figure}

The upcoming Cherenkov Telescope Array (CTA) has the potential to probe the surviving parameter space of these models. Figure \ref{future_prospects:CTA} shows the annihilation cross section, rescaled by the DM fraction (see section \ref{sec:Relic_Abundance} for details), overlayed with projections of the sensitivity after initial construction for the CTA experiment \cite{Acharyya:2020sbj}. The scalar DM model may only be affected at the upper limits of DM masses in the scan, above \(5\)\,TeV. The strongest projected limits will occur for the Dirac fermion model. CTA would seem to have minimal impact on the Majorana model. All of the parameter space that would be constrained corresponds to the off-resonance regions, where \(m_\mathrm{DM} > m_\mathrm{M}\).

We show the predicted number of signal events in DARWIN, a next-generation direct detection experiment \cite{DARWIN}, as a function of the DM mass in Figure \ref{future_prospects:DARWIN}. Each model has the potential to produce tens or even more than a hundred events. DARWIN will therefore prove useful for constraining all the models that we consider here. This should strongly constrain both the scalar and Dirac fermion DM models, including much of the off-resonance regions. For the Majorana model, the DARWIN experiment may impact the lowest DM masses, but much of the high mediator mass regions would remain untouched.

\section{Conclusions}
\label{sec:conclusions}

In this work we performed global scans with \GB of three simplified DM models that interact with quarks via an $s$-channel vector mediator. We found that in the case of scalar DM, a great deal of parameter space with high DM mass and low mediator mass survives, provided that it does not constitute all of the observed relic abundance of DM. Requiring that the model saturates the relic abundance restricts the parameter space to lie either towards the upper mass bounds in this study, or along the region with resonant enhancement in the cross-section. Global constraints on Dirac fermion DM are driven by small excesses in monojet searches, which prefer DM masses of around 260\,GeV and mediator masses of around 540\,GeV. In these regions, the candidate could only constitute a subcomponent of the total relic abundance; requiring that it saturates the DM abundance raises the preferred region to mediator masses of roughly \(1.3\)\,TeV and DM masses of \(580\)\,GeV. Assuming that the collider excesses are only background fluctuations, we also perform scans where the LHC likelihood can only exclude the parameter regions that fit the data worse than the SM. We then see the parameter space open up much more, allowing regions with non-resonant production to survive. Similarly Majorana fermion DM also fits these monojet excesses, and can do so whilst saturating the DM relic abundance. For all of the simplified models that we studied, direct detection, dijet searches and relic abundance measurements provide complementary limits. A lower bound on the DM mass was found for the scalar and Dirac DM models from competing constraints from relic abundance and laboratory experiments, as was found for the EFT in Ref.~\cite{GAMBIT:2021rlp}. This bound is strongly dependent on the scan limits on the couplings. We showed with projections for the CTA and DARWIN experiments that these will be capable of exploring some of the remaining parameter space of the scalar and Dirac fermion models. Also, improvements to dijet limits from Run 3 of the LHC should lower the upper bound on mediator-quark couplings, which will further restrict the surviving parameter space when combined with other constraints.

\section{Acknowledgements}

We thank PRACE for awarding us access to Joliot-Curie at CEA. This project was also undertaken with assistance of resources and services from National Computational Infrastructure, which is supported by the Australian Government. This work was in part performed using the Cambridge Service for Data Driven Discovery (CSD3), part of which is operated by the University of Cambridge Research Computing on behalf of the STFC DiRAC HPC Facility (\url{www.dirac.ac.uk}). The DiRAC component of CSD3 was funded by BEIS capital funding via STFC capital grants ST/P002307/1 and ST/R002452/1 and STFC operations grant ST/R00689X/1. DiRAC is part of the National e-Infrastructure. PS acknowledges funding support from the Australian Research Council under Future Fellowship FT190100814. TEG and FK were funded by the Deutsche Forschungsgemeinschaft (DFG) through the Emmy Noether Grant No. KA 4662/1-1 and grant 396021762 - TRR 257. This article made use of \textsf{pippi v2.2} \cite{pippi}. MJW is supported by the ARC Centre of Excellence for Dark Matter Particle Physics (CE200100008).

\bibliography{R3}

\end{document}

%% file: R3_template.tex
\pdfoutput=1

\usepackage[T1]{fontenc}
\usepackage{lmodern}
\usepackage{calc}
\usepackage{graphicx}
\usepackage{booktabs}
\usepackage{textcomp}
\usepackage{xspace}
\usepackage{relsize}
\usepackage{amssymb}
\usepackage{amsmath}
\usepackage{listings}
\usepackage{microtype}
\usepackage{multirow}
\usepackage{tabularx}
\usepackage{array}
\usepackage{placeins}
\usepackage{cuted}
\usepackage{soul} 
\usepackage{fixltx2e}
\usepackage{slashed}
\usepackage{bm}
\usepackage[numbers,sort&compress]{natbib}
\usepackage[labelfont=bf,font=small]{caption}
\usepackage[skip=-2pt]{subcaption}
\usepackage[clockwise,figuresright]{rotating}
\usepackage{tikz}
\usepackage[normalem]{ulem}
\usepackage[utf8]{inputenc}

\usepackage{etoolbox}
\AfterEndEnvironment{strip}{\leavevmode}

\allowdisplaybreaks

\newcolumntype{L}{>{\raggedright\let\newline\\\arraybackslash\hspace{0pt}}X}
\newcolumntype{R}{>{\raggedleft\let\newline\\\arraybackslash\hspace{0pt}}X}
\newcolumntype{C}{>{\centering\let\newline\\\arraybackslash\hspace{0pt}}X}

\newcommand{\gambitinstitute}[1]{\expandafter\csname #1\endcsname\label{#1}}
\newcommand{\gi}[1]{\gambitinstitute{#1}\and}
\newcommand{\last}[1]{\gambitinstitute{#1}}

\makeatletter

\newcommand{\preprintnumber}[1]{\gdef\@preprintnumber{\begin{flushright}{#1}\end{flushright}}}

\g@addto@macro\bfseries{\boldmath}
\makeatother

\let\underscore\_
\renewcommand{\_}{\discretionary{\underscore}{}{\underscore}}

\makeatletter
\let\orgdescriptionlabel\descriptionlabel
\renewcommand*{\descriptionlabel}[1]{%
  \let\orglabel\label
  \let\label\@gobble
  \phantomsection
  \protected@edef\@currentlabel{#1}%
  \let\label\orglabel
  \orgdescriptionlabel{#1}%
}
\makeatother

\input{jdefs.tex}
\newcommand\cpp[1]{{\lstinline!#1!}}  

\newcommand\yaml[1]{{\lstset{style=yaml}\lstinline!#1!\lstset{style=cpp}}}

\newcommand\term[1]{{\lstset{style=terminal}\lstinline!#1!\lstset{style=cpp}}}
\newcommand\termalt[1]{{\lstset{style=terminalalt}\lstinline!#1!\lstset{style=cpp}}}
\newcommand\fortran[1]{{\lstset{style=fortran}\lstinline!#1!\lstset{style=cpp}}}
\newcommand\py[1]{{\lstset{style=python}\lstinline!#1!\lstset{style=cpp}}}
\newcommand\customtilde{{\raisebox{0.2ex}{\scalebox{0.6}{\boldmath$\sim$}}}}
\newcommand\mathematica[1]{{\lstset{style=Mathematica}\lstinline!#1!\lstset{style=cpp}}}
\newcommand\guminline[1]{{{\lstset{style=gum}\lstinline!#1!}}}
\newcommand\textinline[1]{{{\lstset{style=text}\lstinline!#1!}}}

\def\be{\begin{equation}}
\def\ee{\end{equation}}
\def\ba{\begin{eqnarray}}
\def\ea{\end{eqnarray}}
\newcommand{\bea}{\begin{eqnarray}}
\newcommand{\eea}{\end{eqnarray}}

\lstnewenvironment{lstlistingyaml}{\lstset{style=yaml}}{\lstset{style=cpp}}
\lstnewenvironment{lstlistingterm}{\lstset{style=terminal}}{\lstset{style=cpp}}
\lstnewenvironment{lstlistingfortran}{\lstset{style=fortran}}{\lstset{style=cpp}}
\lstnewenvironment{lstcpp}{\lstset{style=cpp}}{\lstset{style=cpp}}
\lstnewenvironment{lstcppalt}{\lstset{style=cppalt}}{\lstset{style=cpp}}
\lstnewenvironment{lstcppnum}{\lstset{style=cppnum}}{\lstset{style=cpp}}
\lstnewenvironment{lstyaml}{\lstset{style=yaml}}{\lstset{style=cpp}}
\lstnewenvironment{lstgum}{\lstset{style=gum}}{\lstset{style=cpp}}
\lstnewenvironment{lstterm}{\lstset{style=terminal}}{\lstset{style=cpp}}
\lstnewenvironment{lsttermalt}{\lstset{style=terminalalt}}{\lstset{style=cpp}}
\lstnewenvironment{lsttext}{\lstset{style=text}}{\lstset{style=cpp}}
\lstnewenvironment{lstfortran}{\lstset{style=fortran}}{\lstset{style=cpp}}
\lstnewenvironment{lstpy}{\lstset{style=python}}{\lstset{style=cpp}}
\lstnewenvironment{lstmathematica}{\lstset{style=mathematica}}{\lstset{style=cpp}}

\newcommand{\tmpname}{}
\newcommand{\tmplistingname}{}
\makeatletter
\newif\ifATOlabelname
\lst@Key{labelname}{Listing}{\def\ATOlabelname{#1}\global\ATOlabelnametrue}
\makeatother
\lstnewenvironment{lstcpplabel}[1][]{
  \lstset{style=cpp,#1} 
  \ifATOlabelname
    \renewcommand{\tmpname}{\lstlistingname}
    \renewcommand{\tmplistingname}{\lstlistlistingname}
    \renewcommand{\lstlistingname}{\ATOlabelname}
    \renewcommand{\lstlistlistingname}{List of \lstlistingname s}
  \fi
}{
  \renewcommand{\lstlistingname}{\tmpname}
  \renewcommand{\lstlistlistingname}{\tmplistingname}
  \lstset{style=cpp}
}
\definecolor{solarized@base03}{HTML}{002B36}
\definecolor{solarized@base02}{HTML}{073642}
\definecolor{solarized@base01}{HTML}{586e75}
\definecolor{solarized@base00}{HTML}{657b83}
\definecolor{solarized@base0}{HTML}{839496}
\definecolor{solarized@base1}{HTML}{93a1a1}
\definecolor{solarized@base2}{HTML}{EEE8D5}
\definecolor{solarized@base3}{HTML}{FDF6E3}
\definecolor{solarized@yellow}{HTML}{B58900}
\definecolor{solarized@orange}{HTML}{CB4B16}
\definecolor{solarized@red}{HTML}{DC322F}
\definecolor{solarized@magenta}{HTML}{D33682}
\definecolor{solarized@violet}{HTML}{6C71C4}
\definecolor{solarized@blue}{HTML}{268BD2}
\definecolor{solarized@cyan}{HTML}{2AA198}
\definecolor{solarized@green}{HTML}{859900}
\definecolor{darkred}{HTML}{550003}
\definecolor{darkgreen}{HTML}{00AA00}
\definecolor{orchid}{HTML}{AF06F5}

\newcommand\YAMLstringstyle{\footnotesize\color{solarized@green}\mdseries}
\newcommand\YAMLkeystyle{\footnotesize\color{solarized@blue}\ttfamily}
\newcommand\YAMLvaluestyle{\footnotesize\color{blue}\mdseries}
\newcommand\ProcessThreeDashes{\llap{\color{cyan}\mdseries-{-}-}}

\newcommand\CPPcommentstyle{\color{solarized@violet}\footnotesize\ttfamily}
\newcommand\CPPdirectivestyle{\color{solarized@magenta}\footnotesize\ttfamily}
\newcommand\termplainstyle{\footnotesize\ttfamily}

\newcommand\YAMLcommentstyle{\color{solarized@orange}\ttfamily}

\newcommand\processLongMacroDelimiter
{%
\CPPdirectivestyle%
\#define%
}

\lstdefinestyle{cpp}
{
  language=C++,
  basicstyle=\footnotesize\ttfamily,
  basewidth={0.53em,0.44em}, 
  numbers=none,
  tabsize=2,
  breaklines=true,
  escapeinside={@}{@},
  showstringspaces=false,
  numberstyle=\tiny\color{solarized@base01},
  keywordstyle=\color{solarized@orange},
  stringstyle=\color{solarized@red}\ttfamily,
  identifierstyle=\color{solarized@blue},
  commentstyle=\CPPcommentstyle,
  directivestyle=\CPPdirectivestyle,
  emphstyle=\color{solarized@green},
  frame=single,
  rulecolor=\color{solarized@base2},
  rulesepcolor=\color{solarized@base2},
  literate={~} {\customtilde}1,
  moredelim=*[directive]\ \ \#,
  moredelim=*[directive]\ \ \ \ \#
}

\lstdefinestyle{cppalt}
{
  language=C++,
  basicstyle=\footnotesize\ttfamily,
  basewidth={0.53em,0.44em}, 
  numbers=none,
  tabsize=2,
  breaklines=true,
  escapeinside={*@}{@*},
  showstringspaces=false,
  numberstyle=\tiny\color{solarized@base01},
  keywordstyle=\color{solarized@orange},
  stringstyle=\color{solarized@red}\ttfamily,
  identifierstyle=\color{solarized@blue},
  commentstyle=\CPPcommentstyle,
  directivestyle=\CPPdirectivestyle,
  emphstyle=\color{solarized@green},
  frame=single,
  rulecolor=\color{solarized@base2},
  rulesepcolor=\color{solarized@base2},
  literate={~}{\customtilde}1,
  moredelim=**[is][\processLongMacroDelimiter]{BeginLongMacro}{EndLongMacro} 
}

\lstdefinestyle{cppnum}
{
  language=C++,
  basicstyle=\footnotesize\ttfamily,
  basewidth={0.53em,0.44em}, 
  numbers=none,
  tabsize=2,
  breaklines=true,
  escapeinside={@}{@},
  numberstyle=\tiny\color{solarized@base01},
  showstringspaces=false,
  keywordstyle=\color{solarized@orange},
  stringstyle=\color{solarized@red}\ttfamily,
  identifierstyle=\color{solarized@blue},
  commentstyle=\CPPcommentstyle,
  directivestyle=\CPPdirectivestyle,
  emphstyle=\color{solarized@green},
  frame=single,
  rulecolor=\color{solarized@base2},
  rulesepcolor=\color{solarized@base2},
  literate={~} {\customtilde}1,
  moredelim=*[directive]\ \ \#,
  moredelim=*[directive]\ \ \ \ \#
}

\lstdefinestyle{python}
{
  language=Python,
  basicstyle=\footnotesize\ttfamily,
  basewidth={0.53em,0.44em},
  numbers=none,
  tabsize=2,
  breaklines=true,
  escapeinside={@}{@},
  showstringspaces=false,
  numberstyle=\tiny\color{solarized@base01},
  keywordstyle=\color{blue},
  stringstyle=\color{orange}\ttfamily,
  identifierstyle=\color{darkred},
  commentstyle=\color{purple},
  emphstyle=\color{green},
  frame=single,
  rulecolor=\color{solarized@base2},
  rulesepcolor=\color{solarized@base2},
  literate = {~}{\customtilde}1
             {\ as\ }{{\color{blue}\ as\ \color{black}}}3
             {.set}{{\color{black}.}{\color{darkred}set}}4
}

\lstdefinestyle{fortran}
{
  language=Fortran,
  basicstyle=\footnotesize\ttfamily,
  basewidth={0.53em,0.44em},
  numbers=none,
  tabsize=2,
  breaklines=true,
  escapeinside={@}{@},
  showstringspaces=false,
  numberstyle=\tiny\color{solarized@base01},
  keywordstyle=\color{blue},
  stringstyle=\color{orange}\ttfamily,
  identifierstyle=\color{Periwinkle},
  commentstyle=\color{purple},
  emphstyle=\color{green},
  morekeywords={and, or, true, false},
  frame=single,
  rulecolor=\color{solarized@base2},
  rulesepcolor=\color{solarized@base2},
  literate={~}{\customtilde}1
}

\lstdefinestyle{terminal}
{
  language=bash,
  basicstyle=\termplainstyle,
  numbers=none,
  tabsize=2,
  breaklines=true,
  escapeinside={@}{@},
  frame=single,
  showstringspaces=false,
  numberstyle=\tiny\color{solarized@base01},
  keywordstyle=\color{solarized@orange},
  stringstyle=\color{solarized@red}\ttfamily,
  identifierstyle=\color{black},
  commentstyle=\color{solarized@violet},
  emphstyle=\color{solarized@green},
  frame=single,
  rulecolor=\color{solarized@base2},
  rulesepcolor=\color{solarized@base2},
  morekeywords={gambit, cmake, make, mkdir, gum, python, wget, tar, cp, pippi, mpirun},
  deletekeywords={test},
  literate = {/gambit}{{/}{\color{black}}gambit}6
             {gambit/}{{\color{black}}gambit{/}}6
             {gum/}{{\color{black}}gum{/}}4
             {/include}{{/}{\color{black}}include}8
             {cmake/}{{\color{black}}cmake/}6
             {.cmake}{{.}{\color{black}}cmake}6
             {.gum}{{.}{\color{black}}gum}6
             {.tar}{{.}{\color{black}}tar}4
             {source/}{{\color{black}}source{/}}7
             { type}{{\color{black}}{}type}5
             {~}{\customtilde}1
             {math}{{\color{solarized@orange}}math}4
}

\lstdefinestyle{terminalalt}
{
  language=bash,
  basicstyle=\footnotesize\ttfamily,
  numbers=none,
  tabsize=2,
  breaklines=true,
  escapeinside={*@}{@*},
  frame=single,
  showstringspaces=false,
  numberstyle=\tiny\color{solarized@base01},
  keywordstyle=\color{solarized@orange},
  stringstyle=\color{solarized@red}\ttfamily,
  identifierstyle=\color{black},
  commentstyle=\color{solarized@violet},
  emphstyle=\color{solarized@green},
  frame=single,
  rulecolor=\color{solarized@base2},
  rulesepcolor=\color{solarized@base2},
  morekeywords={gambit, cmake, make, mkdir},
  deletekeywords={test},
  literate = {\ gambit}{{\ }{\color{black}}gambit}7
             {/gambit}{{/}{\color{black}}gambit}6
             {gambit/}{{\color{black}}gambit{/}}6
             {/include}{{/}{\color{black}}include}8
             {cmake/}{{\color{black}}cmake/}6
             {.cmake}{{.}{\color{black}}cmake}6
             {~}{\customtilde}1
}

\lstdefinestyle{text}
{
  language={},
  basicstyle=\footnotesize\ttfamily,
  identifierstyle=\color{black},
  numbers=none,
  tabsize=2,
  breaklines=true,
  escapeinside={*@}{@*},
  showstringspaces=false,
  frame=single,
  rulecolor=\color{solarized@base2},
  rulesepcolor=\color{solarized@base2},
  literate={~}{\customtilde}1
}

\lstdefinestyle{yaml}
{
  language=bash,
  escapeinside={@}{@},
  keywords={true,false,null},
  otherkeywords={},
  keywordstyle=\color{solarized@base0}\bfseries,
  basicstyle=\footnotesize\color{black}\ttfamily,
  identifierstyle=\YAMLkeystyle,
  sensitive=false,
  commentstyle=\YAMLcommentstyle,
  morecomment=[l]{\#},
  morecomment=[s]{/*}{*/},
  stringstyle=\YAMLstringstyle\ttfamily,
  moredelim=**[s][\YAMLkeystyle]{,}{:},   
  moredelim=**[l][\YAMLvaluestyle]{:},    
  morestring=[b]',
  morestring=[b]",
  literate =    {---}{{\ProcessThreeDashes}}3
                {>}{{\textcolor{solarized@red}\textgreater}}1
                {gtr}{\textgreater}1
                {grt}{\textgreater}1
                {|}{{\textcolor{solarized@red}\textbar}}1
                {\ -\ }{{\mdseries\color{black}\ -\ \negmedspace}}3
                {\}}{{{\color{black} \}}}}1
                {\{}{{{\color{black} \{}}}1
                {[}{{{\color{black} [}}}1
                {]}{{{\color{black} ]}}}1
                {~}{\customtilde}1,
  breakindent=0pt,
  breakatwhitespace,
  columns=fullflexible
}

\lstdefinestyle{gum}
{
  language=bash,
  escapeinside={@}{@},
  keywords={true,false,null,all},
  otherkeywords={},
  keywordstyle=\color{solarized@base02}\bfseries,
  basicstyle=\footnotesize\color{black}\ttfamily,
  identifierstyle=\color{solarized@magenta},
  sensitive=false,
  commentstyle=\color{solarized@cyan}\ttfamily,
  morecomment=[l]{\#},
  morecomment=[s]{/*}{*/},
  stringstyle=\footnotesize\color{solarized@base01}\mdseries\ttfamily,
  moredelim=**[l][\footnotesize\color{solarized@base02}\mdseries]{:},    
  morestring=[b]',
  morestring=[b]",
  literate =    {---}{{\ProcessThreeDashes}}3
                {grt}{{\textcolor{solarized@magenta}\textgreater}}1
                {gtr}{{\textcolor{solarized@base02}\textgreater}}1
                {/>}{{\textcolor{solarized@magenta}\textgreater}}1
                {/<}{{\textcolor{solarized@magenta}\textless}}1
                {lss}{{\textcolor{solarized@base02}\textless}}1
                {pls}{{\textcolor{solarized@magenta}+}}1
                {mns}{{\textcolor{solarized@magenta}-}}1
                {|}{{\textcolor{solarized@base02}\textbar}}1
                {\ -\ }{{\mdseries\color{solarized@base02}\ -\ \negmedspace}}3
                {\}}{{{\color{solarized@base02} \}}}}1
                {\{}{{{\color{solarized@base02} \{}}}1
                {[}{{{\color{solarized@base02} [}}}1
                {]}{{{\color{solarized@base02} ]}}}1
                {~}{\customtilde}1,
  breakindent=0pt,
  breakatwhitespace,
  columns=fullflexible
}

\lstdefinestyle{mathematica}
{
  language={Mathematica},
  basicstyle=\footnotesize\ttfamily,
  basewidth={0.53em,0.44em},
  numbers=none,
  tabsize=2,
  breaklines=true,
  postbreak=,
  escapeinside={@}{@},
  numberstyle=\tiny\color{black},
  showstringspaces=false,
  numberstyle=\tiny\color{solarized@base01},
  keywordstyle=\color{solarized@orange},
  stringstyle=\color{solarized@red}\ttfamily,
  identifierstyle=\color{solarized@orange}\ttfamily,
  commentstyle=\color{solarized@gray}\ttfamily,
  directivestyle=\color{solarized@orange}\ttfamily,
  emphstyle=\color{solarized@green},
  frame=single,
  rulecolor=\color{solarized@base2},
  rulesepcolor=\color{solarized@base2},
  literate={~} {\customtilde}1,
  moredelim=*[directive]\ \ \#,
  moredelim=*[directive]\ \ \ \ \#,
  mathescape=false
}

\newcommand{\doublecross}[2]{\hyperref[#2]{\textbf{#1}}}
\newcommand{\doublecrosssf}[2]{\hyperref[#2]{\textbf{\textsf{#1}}}}

\newcommand{\startglossary}{\section{Glossary}\label{glossary}Here we explain some terms that have specific technical definitions in \GB.\begin{description}}
\newcommand{\finishglossary}{\end{description}}

\newcommand{\bcode}{\begin{lstlisting}}
\newcommand{\ecode}{\end{lstlisting}}

\newcommand{\gambit}{\textsf{GAMBIT}\xspace}

\newcommand{\darkbit}{\textsf{DarkBit}\xspace}
\newcommand{\colliderbit}{\textsf{ColliderBit}\xspace}

\newcommand{\GB}{\gambit}

\newcommand{\pythia}{\textsf{Pythia}\xspace}

\newcommand{\ds}{\textsf{DarkSUSY}\xspace}
\newcommand{\darksusy}{\ds}

\newcommand\gamLike{\textsf{gamLike}\xspace}
\newcommand\gamlike{\gamLike}

\newcommand\ddcalc{\textsf{DDCalc}\xspace}

\newcommand{\gum}{\textsf{GUM}\xspace}

\newcommand{\fr}{\textsf{FeynRules}\xspace}

\newcommand{\CH}{\textsf{CalcHEP}\xspace}
\newcommand{\MG}{\textsf{MadGraph}\xspace}

\newcommand{\ddm}{\textsf{DirectDM}\xspace}

\newcommand\beq{\begin{equation}}
\newcommand\eeq{\end{equation}}




%% file: S_Channel_Simplified_Models.bbl
\providecommand{\href}[2]{#2}\begingroup\raggedright\begin{thebibliography}{100}

\bibitem{Zwicky33}
F.~{Zwicky}, {\it {Die Rotverschiebung von extragalaktischen Nebeln}},  {\em
  Helvetica Physica Acta} {\bf 6} (1933) 110--127.

\bibitem{bullet}
D.~{Clowe}, M.~{Brada{\v c}}, {\em et.~al.}, {\it {A Direct Empirical Proof of
  the Existence of Dark Matter}},  {\em \apjl} {\bf 648} (2006) L109--L113,
  [\href{http://arxiv.org/abs/astro-ph/0608407}{{\tt astro-ph/0608407}}].

\bibitem{wmap3year}
D.~N. {Spergel}, R.~{Bean}, {\em et.~al.}, {\it {Three-Year Wilkinson Microwave
  Anisotropy Probe (WMAP) Observations: Implications for Cosmology}},  {\em
  \apjs} {\bf 170} (2007) 377--408,
  [\href{http://arxiv.org/abs/astro-ph/0603449}{{\tt astro-ph/0603449}}].

\bibitem{Lee:1977ua}
B.~W. Lee and S.~Weinberg, {\it {Cosmological Lower Bound on Heavy Neutrino
  Masses}},  {\em Phys. Rev. Lett.} {\bf 39} (1977) 165--168.

\bibitem{Arcadi:2017kky}
G.~Arcadi, M.~Dutra, {\em et.~al.}, {\it {The waning of the WIMP? A review of
  models, searches, and constraints}},  {\em Eur. Phys. J.} {\bf C78} (2018)
  203, [\href{http://arxiv.org/abs/1703.07364}{{\tt arXiv:1703.07364}}].

\bibitem{Goodman:2010qn}
J.~Goodman, M.~Ibe, {\em et.~al.}, {\it {Gamma Ray Line Constraints on
  Effective Theories of Dark Matter}},  {\em Nucl. Phys. B} {\bf 844} (2011)
  55--68, [\href{http://arxiv.org/abs/1009.0008}{{\tt arXiv:1009.0008}}].

\bibitem{Beltran:2008xg}
M.~Beltran, D.~Hooper, E.~W. Kolb, and Z.~C. Krusberg, {\it {Deducing the
  nature of dark matter from direct and indirect detection experiments in the
  absence of collider signatures of new physics}},  {\em Phys. Rev. D} {\bf 80}
  (2009) 043509, [\href{http://arxiv.org/abs/0808.3384}{{\tt
  arXiv:0808.3384}}].

\bibitem{Cheung:2011nt}
K.~Cheung, P.-Y. Tseng, and T.-C. Yuan, {\it {Gamma-ray Constraints on
  Effective Interactions of the Dark Matter}},  {\em JCAP} {\bf 06} (2011) 023,
  [\href{http://arxiv.org/abs/1104.5329}{{\tt arXiv:1104.5329}}].

\bibitem{Harnik:2008uu}
R.~Harnik and G.~D. Kribs, {\it {An Effective Theory of Dirac Dark Matter}},
  {\em Phys. Rev. D} {\bf 79} (2009) 095007,
  [\href{http://arxiv.org/abs/0810.5557}{{\tt arXiv:0810.5557}}].

\bibitem{DeSimone:2013gj}
A.~De~Simone, A.~Monin, A.~Thamm, and A.~Urbano, {\it {On the effective
  operators for Dark Matter annihilations}},  {\em JCAP} {\bf 02} (2013) 039,
  [\href{http://arxiv.org/abs/1301.1486}{{\tt arXiv:1301.1486}}].

\bibitem{Karwin:2016tsw}
C.~Karwin, S.~Murgia, T.~M.~P. Tait, T.~A. Porter, and P.~Tanedo, {\it {Dark
  Matter Interpretation of the Fermi-LAT Observation Toward the Galactic
  Center}},  {\em Phys. Rev. D} {\bf 95} (2017) 103005,
  [\href{http://arxiv.org/abs/1612.05687}{{\tt arXiv:1612.05687}}].

\bibitem{Fan:2010gt}
J.~Fan, M.~Reece, and L.-T. Wang, {\it {Non-relativistic effective theory of
  dark matter direct detection}},  {\em \jcap} {\bf 1011} (2010) 042,
  [\href{http://arxiv.org/abs/1008.1591}{{\tt arXiv:1008.1591}}].

\bibitem{Agrawal:2010fh}
P.~Agrawal, Z.~Chacko, C.~Kilic, and R.~K. Mishra, {\it {A Classification of
  Dark Matter Candidates with Primarily Spin-Dependent Interactions with
  Matter}},  \href{http://arxiv.org/abs/1003.1912}{{\tt arXiv:1003.1912}}.

\bibitem{Crivellin:2014gpa}
A.~Crivellin and U.~Haisch, {\it {Dark matter direct detection constraints from
  gauge bosons loops}},  {\em Phys. Rev. D} {\bf 90} (2014) 115011,
  [\href{http://arxiv.org/abs/1408.5046}{{\tt arXiv:1408.5046}}].

\bibitem{Hoferichter:2016nvd}
M.~Hoferichter, P.~Klos, J.~Men\'endez, and A.~Schwenk, {\it {Analysis
  strategies for general spin-independent WIMP-nucleus scattering}},  {\em
  Phys. Rev. D} {\bf 94} (2016) 063505,
  [\href{http://arxiv.org/abs/1605.08043}{{\tt arXiv:1605.08043}}].

\bibitem{Kahlhoefer:2016eds}
F.~Kahlhoefer and S.~Wild, {\it {Studying generalised dark matter interactions
  with extended halo-independent methods}},  {\em JCAP} {\bf 10} (2016) 032,
  [\href{http://arxiv.org/abs/1607.04418}{{\tt arXiv:1607.04418}}].

\bibitem{Buchmueller:2013dya}
O.~Buchmueller, M.~J. Dolan, and C.~McCabe, {\it {Beyond Effective Field Theory
  for Dark Matter Searches at the LHC}},  {\em JHEP} {\bf 01} (2014) 025,
  [\href{http://arxiv.org/abs/1308.6799}{{\tt arXiv:1308.6799}}].

\bibitem{Abdallah:2015ter}
J.~Abdallah {\em et.~al.}, {\it {Simplified Models for Dark Matter Searches at
  the LHC}},  {\em Phys. Dark Univ.} {\bf 9-10} (2015) 8--23,
  [\href{http://arxiv.org/abs/1506.03116}{{\tt arXiv:1506.03116}}].

\bibitem{Abdallah:2014hon}
J.~Abdallah {\em et.~al.}, {\it {Simplified Models for Dark Matter and Missing
  Energy Searches at the LHC}},  \href{http://arxiv.org/abs/1409.2893}{{\tt
  arXiv:1409.2893}}.

\bibitem{Malik:2014ggr}
S.~A. Malik {\em et.~al.}, {\it {Interplay and Characterization of Dark Matter
  Searches at Colliders and in Direct Detection Experiments}},  {\em Phys. Dark
  Univ.} {\bf 9-10} (2015) 51--58, [\href{http://arxiv.org/abs/1409.4075}{{\tt
  arXiv:1409.4075}}].

\bibitem{Busoni:2013lha}
G.~Busoni, A.~De~Simone, E.~Morgante, and A.~Riotto, {\it {On the Validity of
  the Effective Field Theory for Dark Matter Searches at the LHC}},  {\em Phys.
  Lett. B} {\bf 728} (2014) 412--421,
  [\href{http://arxiv.org/abs/1307.2253}{{\tt arXiv:1307.2253}}].

\bibitem{Busoni:2014sya}
G.~Busoni, A.~De~Simone, J.~Gramling, E.~Morgante, and A.~Riotto, {\it {On the
  Validity of the Effective Field Theory for Dark Matter Searches at the LHC,
  Part II: Complete Analysis for the $s$-channel}},  {\em JCAP} {\bf 06} (2014)
  060, [\href{http://arxiv.org/abs/1402.1275}{{\tt arXiv:1402.1275}}].

\bibitem{Busoni:2014haa}
G.~Busoni, A.~De~Simone, T.~Jacques, E.~Morgante, and A.~Riotto, {\it {On the
  Validity of the Effective Field Theory for Dark Matter Searches at the LHC
  Part III: Analysis for the $t$-channel}},  {\em JCAP} {\bf 09} (2014) 022,
  [\href{http://arxiv.org/abs/1405.3101}{{\tt arXiv:1405.3101}}].

\bibitem{GAMBIT:2021rlp}
GAMBIT: P.~Athron {\em et.~al.}, {\it {Thermal WIMPs and the scale of new
  physics: global fits of Dirac dark matter effective field theories}},  {\em
  \epjc} {\bf 81} (2021) 992, [\href{http://arxiv.org/abs/2106.02056}{{\tt
  arXiv:2106.02056}}].

\bibitem{Arina:2018zcq}
C.~Arina, {\it {Impact of cosmological and astrophysical constraints on dark
  matter simplified models}},  {\em Front. Astron. Space Sci.} {\bf 5} (2018)
  30, [\href{http://arxiv.org/abs/1805.04290}{{\tt arXiv:1805.04290}}].

\bibitem{DeSimone:2016fbz}
A.~De~Simone and T.~Jacques, {\it {Simplified models vs. effective field theory
  approaches in dark matter searches}},  {\em \epjc} {\bf 76} (2016) 367,
  [\href{http://arxiv.org/abs/1603.08002}{{\tt arXiv:1603.08002}}].

\bibitem{Albert:2017onk}
A.~Albert {\em et.~al.}, {\it {Recommendations of the LHC Dark Matter Working
  Group: Comparing LHC searches for dark matter mediators in visible and
  invisible decay channels and calculations of the thermal relic density}},
  {\em Phys. Dark Univ.} {\bf 26} (2019) 100377,
  [\href{http://arxiv.org/abs/1703.05703}{{\tt arXiv:1703.05703}}].

\bibitem{Boveia:2016mrp}
A.~Boveia {\em et.~al.}, {\it {Recommendations on presenting LHC searches for
  missing transverse energy signals using simplified $s$-channel models of dark
  matter}},  {\em Phys. Dark Univ.} {\bf 27} (2020) 100365,
  [\href{http://arxiv.org/abs/1603.04156}{{\tt arXiv:1603.04156}}].

\bibitem{Kahlhoefer:2017dnp}
F.~Kahlhoefer, {\it {Review of LHC Dark Matter Searches}},  {\em Int. J. Mod.
  Phys. A} {\bf 32} (2017) 1730006,
  [\href{http://arxiv.org/abs/1702.02430}{{\tt arXiv:1702.02430}}].

\bibitem{Morgante:2018tiq}
E.~Morgante, {\it {Simplified Dark Matter Models}},  {\em Adv. High Energy
  Phys.} {\bf 2018} (2018) 5012043,
  [\href{http://arxiv.org/abs/1804.01245}{{\tt arXiv:1804.01245}}].

\bibitem{DEramo:2016gos}
F.~D'Eramo, B.~J. Kavanagh, and P.~Panci, {\it {You can hide but you have to
  run: direct detection with vector mediators}},  {\em JHEP} {\bf 08} (2016)
  111, [\href{http://arxiv.org/abs/1605.04917}{{\tt arXiv:1605.04917}}].

\bibitem{Carpenter:2016thc}
L.~M. Carpenter, R.~Colburn, J.~Goodman, and T.~Linden, {\it {Indirect
  Detection Constraints on s and t Channel Simplified Models of Dark Matter}},
  {\em Phys. Rev. D} {\bf 94} (2016) 055027,
  [\href{http://arxiv.org/abs/1606.04138}{{\tt arXiv:1606.04138}}].

\bibitem{Abercrombie:2015wmb}
D.~Abercrombie {\em et.~al.}, {\it {Dark Matter Benchmark Models for Early LHC
  Run-2 Searches: Report of the ATLAS/CMS Dark Matter Forum}},  {\em Phys. Dark
  Univ.} {\bf 26} (2019) 100371, [\href{http://arxiv.org/abs/1507.00966}{{\tt
  arXiv:1507.00966}}].

\bibitem{Bagnaschi:2019djj}
E.~Bagnaschi {\em et.~al.}, {\it {Global Analysis of Dark Matter Simplified
  Models with Leptophobic Spin-One Mediators using MasterCode}},  {\em \epjc}
  {\bf 79} (2019) 895, [\href{http://arxiv.org/abs/1905.00892}{{\tt
  arXiv:1905.00892}}].

\bibitem{gambit}
\GB Collaboration: P.~{Athron}, C.~{Bal{\'a}zs}, {\em et.~al.}, {\it {GAMBIT:
  The Global and Modular Beyond-the-Standard-Model Inference Tool}},  {\em
  \epjc} {\bf 77} (2017) 784, [\href{http://arxiv.org/abs/1705.07908}{{\tt
  arXiv:1705.07908}}]. Addendum in \cite{gambit_addendum}.

\bibitem{GUM}
\GB Collaboration: S.~Bloor, T.~E. Gonzalo, {\em et.~al.}, {\it {The GAMBIT
  Universal Model Machine: from Lagrangians to likelihoods}},  {\em \epjc} {\bf
  81} (2021) 1103, [\href{http://arxiv.org/abs/2107.00030}{{\tt
  arXiv:2107.00030}}].

\bibitem{Zenodo_DMsimp}
\GB Collaboration, \textit{Supplementary Data: Global Fits of simplified models
  for dark matter with \GB. I. Scalar and fermionic models with s-channel
  vector mediators.}, (2022),
  \href{https://zenodo.org/record/6615830}{\nolinkurl{https://zenodo.org/record/6615830}}.

\bibitem{DMSimpII}
C.~Chang, P.~Scott, T.~E. Gonzalo, F.~Kahlhoefer, and M.~White, {\it {Global
  fits of simplified models for dark matter with GAMBIT II. Vector dark matter
  with an $s$-channel vector mediator}},  {\em Eur. Phys. J. C} {\bf 83} (2023)
  692, [\href{http://arxiv.org/abs/2303.08351}{{\tt arXiv:2303.08351}}].

\bibitem{Duerr:2016tmh}
M.~Duerr, F.~Kahlhoefer, K.~Schmidt-Hoberg, T.~Schwetz, and S.~Vogl, {\it {How
  to save the WIMP: global analysis of a dark matter model with two s-channel
  mediators}},  {\em JHEP} {\bf 09} (2016) 042,
  [\href{http://arxiv.org/abs/1606.07609}{{\tt arXiv:1606.07609}}].

\bibitem{Duerr:2014wra}
M.~Duerr and P.~Fileviez~Perez, {\it {Theory for Baryon Number and Dark Matter
  at the LHC}},  {\em Phys. Rev. D} {\bf 91} (2015) 095001,
  [\href{http://arxiv.org/abs/1409.8165}{{\tt arXiv:1409.8165}}].

\bibitem{Ellis:2017tkh}
J.~Ellis, M.~Fairbairn, and P.~Tunney, {\it {Anomaly-Free Dark Matter Models
  are not so Simple}},  {\em JHEP} {\bf 08} (2017) 053,
  [\href{http://arxiv.org/abs/1704.03850}{{\tt arXiv:1704.03850}}].

\bibitem{Kahlhoefer:2015bea}
F.~Kahlhoefer, K.~Schmidt-Hoberg, T.~Schwetz, and S.~Vogl, {\it {Implications
  of unitarity and gauge invariance for simplified dark matter models}},  {\em
  JHEP} {\bf 02} (2016) 016, [\href{http://arxiv.org/abs/1510.02110}{{\tt
  arXiv:1510.02110}}].

\bibitem{Boehm:2003hm}
C.~Boehm and P.~Fayet, {\it {Scalar dark matter candidates}},  {\em Nucl. Phys.
  B} {\bf 683} (2004) 219--263,
  [\href{http://arxiv.org/abs/hep-ph/0305261}{{\tt hep-ph/0305261}}].

\bibitem{Agnese:2015nto}
SuperCDMS: R.~Agnese {\em et.~al.}, {\it {New Results from the Search for
  Low-Mass Weakly Interacting Massive Particles with the CDMS Low Ionization
  Threshold Experiment}},  {\em \prl} {\bf 116} (2016) 071301,
  [\href{http://arxiv.org/abs/1509.02448}{{\tt arXiv:1509.02448}}].

\bibitem{Angloher:2015ewa}
CRESST: G.~Angloher {\em et.~al.}, {\it {Results on light dark matter particles
  with a low-threshold CRESST-II detector}},  {\em Eur. Phys. J.} {\bf C76}
  (2016) 25, [\href{http://arxiv.org/abs/1509.01515}{{\tt arXiv:1509.01515}}].

\bibitem{Abdelhameed:2019hmk}
CRESST: A.~H. Abdelhameed {\em et.~al.}, {\it {First results from the
  CRESST-III low-mass dark matter program}},  {\em \prd} {\bf 100} (2019)
  102002, [\href{http://arxiv.org/abs/1904.00498}{{\tt arXiv:1904.00498}}].

\bibitem{Agnes:2018fwg}
DarkSide: P.~Agnes {\em et.~al.}, {\it {DarkSide-50 532-day Dark Matter Search
  with Low-Radioactivity Argon}},  {\em \prd} {\bf 98} (2018) 102006,
  [\href{http://arxiv.org/abs/1802.07198}{{\tt arXiv:1802.07198}}].

\bibitem{LUXrun2}
LUX: D.~S. Akerib {\em et.~al.}, {\it {Results from a search for dark matter in
  the complete LUX exposure}},  {\em \prl} {\bf 118} (2017) 021303,
  [\href{http://arxiv.org/abs/1608.07648}{{\tt arXiv:1608.07648}}].

\bibitem{Amole:2017dex}
PICO: C.~Amole {\em et.~al.}, {\it {Dark Matter Search Results from the PICO-60
  C$_3$F$_8$ Bubble Chamber}},  {\em \prl} {\bf 118} (2017) 251301,
  [\href{http://arxiv.org/abs/1702.07666}{{\tt arXiv:1702.07666}}].

\bibitem{Amole:2019fdf}
PICO: C.~Amole {\em et.~al.}, {\it {Dark Matter Search Results from the
  Complete Exposure of the PICO-60 C$_3$F$_8$ Bubble Chamber}},  {\em \prd}
  {\bf 100} (2019) 022001, [\href{http://arxiv.org/abs/1902.04031}{{\tt
  arXiv:1902.04031}}].

\bibitem{Tan:2016zwf}
PandaX-II: A.~Tan {\em et.~al.}, {\it {Dark Matter Results from First 98.7 Days
  of Data from the PandaX-II Experiment}},  {\em \prl} {\bf 117} (2016) 121303,
  [\href{http://arxiv.org/abs/1607.07400}{{\tt arXiv:1607.07400}}].

\bibitem{Cui:2017nnn}
PandaX-II: X.~Cui {\em et.~al.}, {\it {Dark Matter Results From 54-Ton-Day
  Exposure of PandaX-II Experiment}},  {\em \prl} {\bf 119} (2017) 181302,
  [\href{http://arxiv.org/abs/1708.06917}{{\tt arXiv:1708.06917}}].

\bibitem{Aprile:2018dbl}
XENON: E.~Aprile {\em et.~al.}, {\it {Dark Matter Search Results from a One
  Ton-Year Exposure of XENON1T}},  {\em \prl} {\bf 121} (2018) 111302,
  [\href{http://arxiv.org/abs/1805.12562}{{\tt arXiv:1805.12562}}].

\bibitem{LZ:2022ufs}
LZ: J.~Aalbers {\em et.~al.}, {\it {First Dark Matter Search Results from the
  LUX-ZEPLIN (LZ) Experiment}},  \href{http://arxiv.org/abs/2207.03764}{{\tt
  arXiv:2207.03764}}.

\bibitem{CMS:2019gwf}
CMS: A.~M. Sirunyan {\em et.~al.}, {\it {Search for high mass dijet resonances
  with a new background prediction method in proton-proton collisions at
  $\sqrt{s} =$ 13 TeV}},  {\em JHEP} {\bf 05} (2020) 033,
  [\href{http://arxiv.org/abs/1911.03947}{{\tt arXiv:1911.03947}}].

\bibitem{ATLAS:2019fgd}
ATLAS: G.~Aad {\em et.~al.}, {\it {Search for new resonances in mass
  distributions of jet pairs using 139 fb$^{-1}$ of $pp$ collisions at
  $\sqrt{s}=13$ TeV with the ATLAS detector}},  {\em JHEP} {\bf 03} (2020) 145,
  [\href{http://arxiv.org/abs/1910.08447}{{\tt arXiv:1910.08447}}].

\bibitem{ATLAS:2018qto}
ATLAS: M.~Aaboud {\em et.~al.}, {\it {Search for low-mass dijet resonances
  using trigger-level jets with the ATLAS detector in $pp$ collisions at
  $\sqrt{s}=13$ TeV}},  {\em Phys. Rev. Lett.} {\bf 121} (2018) 081801,
  [\href{http://arxiv.org/abs/1804.03496}{{\tt arXiv:1804.03496}}].

\bibitem{CDF:2008ieg}
CDF: T.~Aaltonen {\em et.~al.}, {\it {Search for new particles decaying into
  dijets in proton-antiproton collisions at $\sqrt{s}=1.96$ TeV}},  {\em Phys.
  Rev. D} {\bf 79} (2009) 112002, [\href{http://arxiv.org/abs/0812.4036}{{\tt
  arXiv:0812.4036}}].

\bibitem{ATLAS:2018hbc}
ATLAS: M.~Aaboud {\em et.~al.}, {\it {Search for light resonances decaying to
  boosted quark pairs and produced in association with a photon or a jet in
  proton-proton collisions at $\sqrt{s}=13$ TeV with the ATLAS detector}},
  {\em Phys. Lett. B} {\bf 788} (2019) 316--335,
  [\href{http://arxiv.org/abs/1801.08769}{{\tt arXiv:1801.08769}}].

\bibitem{ATLAS:2018hzj}
ATLAS Collaboration, {\it {Search for boosted resonances decaying to two
  b-quarks and produced in association with a jet at $\sqrt{s}=13$ TeV with the
  ATLAS detector}},  2018.

\bibitem{CMS:2019emo}
CMS: A.~M. Sirunyan {\em et.~al.}, {\it {Search for low mass vector resonances
  decaying into quark-antiquark pairs in proton-proton collisions at
  $\sqrt{s}=$ 13 TeV}},  {\em Phys. Rev. D} {\bf 100} (2019) 112007,
  [\href{http://arxiv.org/abs/1909.04114}{{\tt arXiv:1909.04114}}].

\bibitem{ATLAS:2019itm}
ATLAS: M.~Aaboud {\em et.~al.}, {\it {Search for low-mass resonances decaying
  into two jets and produced in association with a photon using $pp$ collisions
  at $\sqrt{s} = 13$ TeV with the ATLAS detector}},  {\em Phys. Lett. B} {\bf
  795} (2019) 56--75, [\href{http://arxiv.org/abs/1901.10917}{{\tt
  arXiv:1901.10917}}].

\bibitem{CMS:2019xai}
CMS: A.~M. Sirunyan {\em et.~al.}, {\it {Search for Low-Mass Quark-Antiquark
  Resonances Produced in Association with a Photon at $\sqrt {s}$ =13 TeV}},
  {\em Phys. Rev. Lett.} {\bf 123} (2019) 231803,
  [\href{http://arxiv.org/abs/1905.10331}{{\tt arXiv:1905.10331}}].

\bibitem{Aad:2021egl}
ATLAS: G.~Aad {\em et.~al.}, {\it {Search for new phenomena in events with an
  energetic jet and missing transverse momentum in $pp$ collisions at $\sqrt{s}
  = 13$ TeV with the ATLAS detector}},
  \href{http://arxiv.org/abs/2102.10874}{{\tt arXiv:2102.10874}}.

\bibitem{CMS:2021snz}
CMS collaboration,{\it { Search for new particles in events with energetic jets
  and large missing transverse momentum in proton-proton collisions at
  $\sqrt{s}=13~\mathrm{TeV}$}},  {\em CMS-PAS-EXO-20-004} (2021).

\bibitem{LATdwarfP8}
Fermi-LAT: M.~Ackermann {\em et.~al.}, {\it {Searching for Dark Matter
  Annihilation from Milky Way Dwarf Spheroidal Galaxies with Six Years of Fermi
  Large Area Telescope Data}},  {\em \prl} {\bf 115} (2015) 231301,
  [\href{http://arxiv.org/abs/1503.02641}{{\tt arXiv:1503.02641}}].

\bibitem{Aghanim:2018eyx}
Planck: N.~Aghanim {\em et.~al.}, {\it {Planck 2018 results. VI. Cosmological
  parameters}},  {\em Astron. Astrophys.} {\bf 641} (2020) A6,
  [\href{http://arxiv.org/abs/1807.06209}{{\tt arXiv:1807.06209}}].

\bibitem{Gondolo:1990dk}
P.~Gondolo and G.~Gelmini, {\it {Cosmic abundances of stable particles:
  Improved analysis}},  {\em \nphysa} {\bf 360} (1991) 145--179.

\bibitem{Binder:2017rgn}
T.~Binder, T.~Bringmann, M.~Gustafsson, and A.~Hryczuk, {\it {Early kinetic
  decoupling of dark matter: when the standard way of calculating the thermal
  relic density fails}},  {\em \prd} {\bf 96} (2017) 115010,
  [\href{http://arxiv.org/abs/1706.07433}{{\tt arXiv:1706.07433}}].

\bibitem{Kaplan:2009ag}
D.~E. Kaplan, M.~A. Luty, and K.~M. Zurek, {\it {Asymmetric Dark Matter}},
  {\em \prd} {\bf 79} (2009) 115016,
  [\href{http://arxiv.org/abs/0901.4117}{{\tt arXiv:0901.4117}}].

\bibitem{Pukhov:2004ca}
A.~Pukhov, {\it {CalcHEP 2.3: MSSM, structure functions, event generation,
  batchs, and generation of matrix elements for other packages}},
  \href{http://arxiv.org/abs/hep-ph/0412191}{{\tt hep-ph/0412191}}.

\bibitem{Belyaev:2012qa}
A.~Belyaev, N.~D. Christensen, and A.~Pukhov, {\it {CalcHEP 3.4 for collider
  physics within and beyond the Standard Model}},  {\em \cpc} {\bf 184} (2013)
  1729--1769, [\href{http://arxiv.org/abs/1207.6082}{{\tt arXiv:1207.6082}}].

\bibitem{darksusy}
T.~Bringmann, J.~Edsjö, P.~Gondolo, P.~Ullio, and L.~Bergström, {\it
  {DarkSUSY 6 : An Advanced Tool to Compute Dark Matter Properties
  Numerically}},  {\em \jcap} {\bf 1807} (2018) 033,
  [\href{http://arxiv.org/abs/1802.03399}{{\tt arXiv:1802.03399}}].

\bibitem{darksusy4}
P.~Gondolo, J.~Edsjo, {\em et.~al.}, {\it {DarkSUSY: Computing supersymmetric
  dark matter properties numerically}},  {\em \jcap} {\bf 0407} (2004) 008,
  [\href{http://arxiv.org/abs/astro-ph/0406204}{{\tt astro-ph/0406204}}].

\bibitem{Arbey:2021gdg}
A.~Arbey and F.~Mahmoudi, {\it {Dark matter and the early Universe: A review}},
   {\em Prog. Part. Nucl. Phys.} {\bf 119} (2021) 103865,
  [\href{http://arxiv.org/abs/2104.11488}{{\tt arXiv:2104.11488}}].

\bibitem{DarkBit}
\GB Dark Matter Workgroup: T.~{Bringmann}, J.~{Conrad}, {\em et.~al.}, {\it
  {DarkBit: A GAMBIT module for computing dark matter observables and
  likelihoods}},  {\em \epjc} {\bf 77} (2017) 831,
  [\href{http://arxiv.org/abs/1705.07920}{{\tt arXiv:1705.07920}}].

\bibitem{Fitzpatrick:2012ix}
A.~L. Fitzpatrick, W.~Haxton, E.~Katz, N.~Lubbers, and Y.~Xu, {\it {The
  Effective Field Theory of Dark Matter Direct Detection}},  {\em \jcap} {\bf
  1302} (2013) 004, [\href{http://arxiv.org/abs/1203.3542}{{\tt
  arXiv:1203.3542}}].

\bibitem{Anand:2013yka}
N.~Anand, A.~L. Fitzpatrick, and W.~C. Haxton, {\it {Weakly interacting massive
  particle-nucleus elastic scattering response}},  {\em \prc} {\bf 89} (2014)
  065501, [\href{http://arxiv.org/abs/1308.6288}{{\tt arXiv:1308.6288}}].

\bibitem{Dent:2015zpa}
J.~B. Dent, L.~M. Krauss, J.~L. Newstead, and S.~Sabharwal, {\it {General
  analysis of direct dark matter detection: From microphysics to observational
  signatures}},  {\em \prd} {\bf 92} (2015) 063515,
  [\href{http://arxiv.org/abs/1505.03117}{{\tt arXiv:1505.03117}}].

\bibitem{HP}
GAMBIT Collaboration: P.~Athron {\em et.~al.}, {\it {Global analyses of Higgs
  portal singlet dark matter models using GAMBIT}},  {\em \epjc} {\bf 79}
  (2019) 38, [\href{http://arxiv.org/abs/1808.10465}{{\tt arXiv:1808.10465}}].

\bibitem{Baum:2017kfa}
S.~Baum, R.~Catena, J.~Conrad, K.~Freese, and M.~B. Krauss, {\it {Determining
  dark matter properties with a XENONnT/LZ signal and LHC Run 3 monojet
  searches}},  {\em Phys. Rev. D} {\bf 97} (2018) 083002,
  [\href{http://arxiv.org/abs/1709.06051}{{\tt arXiv:1709.06051}}].

\bibitem{Bringmann:2012ez}
T.~Bringmann and C.~Weniger, {\it {Gamma Ray Signals from Dark Matter:
  Concepts, Status and Prospects}},  {\em Phys. Dark Univ.} {\bf 1} (2012)
  194--217, [\href{http://arxiv.org/abs/1208.5481}{{\tt arXiv:1208.5481}}].

\bibitem{Acharyya:2020sbj}
CTA: A.~Acharyya {\em et.~al.}, {\it {Sensitivity of the Cherenkov Telescope
  Array to a dark matter signal from the Galactic centre}},  {\em JCAP} {\bf
  01} (2021) 057, [\href{http://arxiv.org/abs/2007.16129}{{\tt
  arXiv:2007.16129}}].

\bibitem{Bauer:2017fsw}
M.~Bauer, M.~Klassen, and V.~Tenorth, {\it {Universal properties of
  pseudoscalar mediators in dark matter extensions of 2HDMs}},  {\em JHEP} {\bf
  07} (2018) 107, [\href{http://arxiv.org/abs/1712.06597}{{\tt
  arXiv:1712.06597}}].

\bibitem{Zhou:2013fla}
N.~Zhou, D.~Berge, and D.~Whiteson, {\it {Mono-everything: combined limits on
  dark matter production at colliders from multiple final states}},  {\em Phys.
  Rev. D} {\bf 87} (2013) 095013, [\href{http://arxiv.org/abs/1302.3619}{{\tt
  arXiv:1302.3619}}].

\bibitem{Brennan:2016xjh}
A.~J. Brennan, M.~F. McDonald, J.~Gramling, and T.~D. Jacques, {\it {Collide
  and Conquer: Constraints on Simplified Dark Matter Models using Mono-X
  Collider Searches}},  {\em JHEP} {\bf 05} (2016) 112,
  [\href{http://arxiv.org/abs/1603.01366}{{\tt arXiv:1603.01366}}].

\bibitem{ColliderBit}
\GB Collider Workgroup: C.~{Bal{\'a}zs}, A.~{Buckley}, {\em et.~al.}, {\it
  {ColliderBit: a GAMBIT module for the calculation of high-energy collider
  observables and likelihoods}},  {\em \epjc} {\bf 77} (2017) 795,
  [\href{http://arxiv.org/abs/1705.07919}{{\tt arXiv:1705.07919}}].

\bibitem{Buckley:2014fba}
M.~R. Buckley, D.~Feld, and D.~Goncalves, {\it {Scalar Simplified Models for
  Dark Matter}},  {\em \prd} {\bf 91} (2015) 015017,
  [\href{http://arxiv.org/abs/1410.6497}{{\tt arXiv:1410.6497}}].

\bibitem{Alwall:2011uj}
J.~Alwall, M.~Herquet, F.~Maltoni, O.~Mattelaer, and T.~Stelzer, {\it {MadGraph
  5 : Going Beyond}},  {\em \jhep} {\bf 06} (2011) 128,
  [\href{http://arxiv.org/abs/1106.0522}{{\tt arXiv:1106.0522}}].

\bibitem{Sjostrand:2007gs}
T.~Sjostrand, S.~Mrenna, and P.~Z. Skands, {\it {A Brief Introduction to PYTHIA
  8.1}},  {\em Comput. Phys. Commun.} {\bf 178} (2008) 852--867,
  [\href{http://arxiv.org/abs/0710.3820}{{\tt arXiv:0710.3820}}].

\bibitem{Degrande:2011ua}
C.~Degrande, C.~Duhr, {\em et.~al.}, {\it {UFO - The Universal FeynRules
  Output}},  {\em Comput. Phys. Commun.} {\bf 183} (2012) 1201--1214,
  [\href{http://arxiv.org/abs/1108.2040}{{\tt arXiv:1108.2040}}].

\bibitem{Alloul:2013bka}
A.~Alloul, N.~D. Christensen, C.~Degrande, C.~Duhr, and B.~Fuks, {\it
  {FeynRules 2.0 - A complete toolbox for tree-level phenomenology}},  {\em
  Comput. Phys. Commun.} {\bf 185} (2014) 2250--2300,
  [\href{http://arxiv.org/abs/1310.1921}{{\tt arXiv:1310.1921}}].

\bibitem{Conte:2012fm}
E.~Conte, B.~Fuks, and G.~Serret, {\it {MadAnalysis 5, A User-Friendly
  Framework for Collider Phenomenology}},  {\em Comput. Phys. Commun.} {\bf
  184} (2013) 222--256, [\href{http://arxiv.org/abs/1206.1599}{{\tt
  arXiv:1206.1599}}].

\bibitem{Collaboration:2242860}
CMS Collaboration, {\it {Simplified likelihood for the re-interpretation of
  public CMS results}},   CMS-NOTE-2017-001, 2017.

\bibitem{ATL-PHYS-PUB-2019-029}
ATLAS Collaboration, {\it {Reproducing searches for new physics with the ATLAS
  experiment through publication of full statistical likelihoods}},  2019.

\bibitem{EWMSSM}
GAMBIT Collaboration: P.~Athron {\em et.~al.}, {\it {Combined collider
  constraints on neutralinos and charginos}},  {\em \epjc} {\bf 79} (2019) 395,
  [\href{http://arxiv.org/abs/1809.02097}{{\tt arXiv:1809.02097}}].

\bibitem{DMEFT}
GAMBIT: P.~Athron {\em et.~al.}, {\it {Thermal WIMPs and the scale of new
  physics: global fits of Dirac dark matter effective field theories}},  {\em
  \epjc} {\bf 81} (2021) 992, [\href{http://arxiv.org/abs/2106.02056}{{\tt
  arXiv:2106.02056}}].

\bibitem{Reid:2014boa}
M.~J. Reid {\em et.~al.}, {\it {Trigonometric Parallaxes of High Mass Star
  Forming Regions: the Structure and Kinematics of the Milky Way}},  {\em
  Astrophys. J.} {\bf 783} (2014) 130,
  [\href{http://arxiv.org/abs/1401.5377}{{\tt arXiv:1401.5377}}].

\bibitem{Deason:2019kgj}
A.~J. Deason, A.~Fattahi, {\em et.~al.}, {\it The local high-velocity tail and
  the galactic escape speed},  {\em \mnras} {\bf 485} (2019) 3514–3526,
  [\href{http://arxiv.org/abs/1901.02016}{{\tt arXiv:1901.02016}}].

\bibitem{ScannerBit}
\GB Scanner Workgroup: G.~D. {Martinez}, J.~{McKay}, {\em et.~al.}, {\it
  {Comparison of statistical sampling methods with ScannerBit, the GAMBIT
  scanning module}},  {\em \epjc} {\bf 77} (2017) 761,
  [\href{http://arxiv.org/abs/1705.07959}{{\tt arXiv:1705.07959}}].

\bibitem{DARWIN}
DARWIN: J.~Aalbers {\em et.~al.}, {\it {DARWIN: towards the ultimate dark
  matter detector}},  {\em \jcap} {\bf 1611} (2016) 017,
  [\href{http://arxiv.org/abs/1606.07001}{{\tt arXiv:1606.07001}}].

\bibitem{pippi}
P.~{Scott}, {\it {Pippi -- painless parsing, post-processing and plotting of
  posterior and likelihood samples}},  {\em \epjp} {\bf 127} (2012) 138,
  [\href{http://arxiv.org/abs/1206.2245}{{\tt arXiv:1206.2245}}].

\bibitem{gambit_addendum}
\GB Collaboration: P.~{Athron}, C.~{Bal{\'a}zs}, {\em et.~al.}, {\it {GAMBIT:
  The Global and Modular Beyond-the-Standard-Model Inference Tool. Addendum for
  GAMBIT 1.1: Mathematica backends, SUSYHD interface and updated likelihoods}},
   {\em \epjc} {\bf 78} (2018) 98, [\href{http://arxiv.org/abs/1705.07908}{{\tt
  arXiv:1705.07908}}]. Addendum to \cite{gambit}.

\end{thebibliography}\endgroup
